\documentstyle[11pt]{article}
\author {Wen-Xiu Ma\thanks{On leave of absence from
 Institute of Mathematics, Fudan University, 
Shanghai 200433, P. R. of China;
Current Fax: 0049-5251-603836 and Email: wenxiuma@uni-paderborn.de}
\ and Benno Fuchssteiner\\
FB Mathematik-Informatik, Universit\"at Paderborn,
\\ D-33098 Paderborn, Germany}
\title
{ Integrable Theory of the Perturbation Equation}
\date{\nonumber}
\begin{document}

\setlength{\baselineskip}{15.5pt}
\maketitle
\begin{abstract}
An integrable theory is developed for the perturbation equations engendered
from small disturbances of solutions. 
It includes various integrable properties
of the perturbation equations:
hereditary recursion operators,
master symmetries,
linear representations (Lax and zero curvature representations) and 
Hamiltonian structures etc. and provides
 us a method to generate hereditary operators, Hamiltonian
operators and symplectic operators starting from the known ones.
The resulting perturbation equations give rise to a sort of 
 integrable coupling
of soliton equations. 
Two examples (MKdV hierarchy and KP equation) are 
carefully carried out.
\end{abstract}


\newcommand{\R}{\mbox{\rm I \hspace{-0.9em} R}}
\def \3 {\hat{u}_N}
\def \4 {\hat{\eta}_N}
\def \5 {\hat{K}_N}
\def \6 {\hat{S}_N}
\def \7 {\hat{J}_N}
\def \P {\hat{\Phi}_N}
\def \J {\hat{J}_N}
\def \la {\lambda}
\def \1 {\eta}
\def \2 {\varepsilon}
\def \La {\Lambda}
\def \be {\beta}
\def \al{\alpha}
\def \del{\delta}
\def \Del{\Delta}
\def \al{\alpha}
\def \vare{\varepsilon}

\def \part {\partial}
\def \h1 \hat{\eta }
\def \be {\begin{equation}}
\def \ee {\end{equation}}
\def \ba {\begin{array}}
\def \ea {\end{array}}

\newcommand{\eqnsection}{
   \renewcommand{\theequation}{\thesection.\arabic{equation}}
   \makeatletter
   \csname $addtoreset\endcsname
   \makeatother}
\eqnsection

\newtheorem{lemma}{Lemma}[section]
\newtheorem{theorem}{Theorem}[section]
\newtheorem{definition}{Definition}[section]

\section{Introduction}
\setcounter{equation}{0}

Integrable nonlinear wave or evolution equations (for instance,
KdV, NLS, SG and KP equations) are ideal mathematical models of
real physical phenomena although they play an outstanding role in 
physical problems. Therefore for these equations
we often need to take into account 
the effect of small perturbation so that their applicability may be 
extended to higher order nonlinearity or larger amplitude waves.
There are mainly two kinds of perturbation 
worthy studying for integrable equations.
The one is the perturbation situation
 of integrable equations
themselves and the other one, 
the perturbation situation of solutions of the original unperturbed 
integrable equations. They all can provide approximate solutions
to real physical problems.
 
In the context of soliton perturbation,
what one considers usually is the first kind of  perturbation situation.
There have been quiet a few of powerful techniques for dealing with 
this kind of the perturbation (see \cite{Karpman}\, \cite{KeenerMacLaughlin}
\cite{KodamaAblowitz}\, \cite{Olver2}\,
\cite{Herman} and references therein). Various perturbed cases
of integrable equations have been considered, including 
the perturbed KdV and MKdV equations$^{\cite{KodamaAblowitz}\,
\cite{Tanaka}\,
\cite{GrimshawMitsudera}}$, 
the perturbed nonlinear Schr\"odinger equation$^{\cite{ElginBJ}\,
\cite{BallantyneGT}}$, 
the perturbed Burgers equation$^{\cite{CurroDP}}$ and the perturbed 
Benjamin-Ono equation$^{\cite{Matsuno}}$ etc. 
A detailed survey for this kind of perturbation theory based upon the inverse 
scattering transformation was provided by Kivshar and Malomed 
\cite{KivsharMalomed}.
However there don't exist
 so many works devoted to the second kind of perturbation
situation. Among them there are the following several
works. Tamizhmani and Lakshmanan considered the complete integrability
of the perturbed equations by small disturbance of solutions of KdV 
equation$^{\cite{TamizhmaniLakshmanan}}$.
We analyzed the similar perturbation 
situation of the whole KdV integrable hierarchy
and pointed out a mistake in Ref. \cite{TamizhmaniLakshmanan} for the 
Hamiltonian 
structure$^{\cite{MaFuchsstener}}$. 
Recently it has been observed by Kraenkel et al. that a multiple
time scale expansion may relate the 
solutions of long surface water-waves and the Boussinesq
equation to KdV integrable hierarchy$^{\cite{KraenkelMP}\,\cite{KraenkelMMP}}$.

In this paper we would like to consider the second kind of perturbation.
Mathematically, this kind of perturbation yields 
interesting results. For example, we shall show that
it preserves complete integrability. 
In other words, the equations generated by the perturbation
are still integrable and thus give rise to new examples of integrable 
equations. Moreover they are all special integrable coupling of the original
integrable equations. 

We now introduce our notation and conception. Some notation comes from
Refs. \cite{FuchssteinerFokas}, 
\cite{Fuchssteiner1}, \cite{Oevel}. 
Let $M=M(u)$ be a suitable manifold possessing a manifold variable $u$ 
(we write $u$ as a column vector) and 
$\hat {M}_N=\hat {M}_N(\hat {\1 }_N)$ 
be another  suitable manifold possessing a manifold variable 
$\hat {\1 }_N =(\1 _0^T,\1 _1^T,\cdots,\1 _N^T)^T,$ $ N\ge1$, 
where $\eta _i$, $0\le i\le N,$ are column vectors and $T$ means 
the transpose of matrices. Assume that 
$T(M), T(\hat{M}_N)$ denote the tangent bundles on $M$ and $\hat{M}_N$,
 $T^*(M), T^*(\hat{M}_N)$
denote the cotangent bundles
 on $M$ and $\hat{M}_N$, and $C^\infty (M)$, $C^\infty (\hat{M}_N)$
denote the spaces of smooth functions on $M$ and $ \hat{M}_N$, respectively.
Throughout this paper we require that the column vector 
$\eta _i$ $(0\le i\le N)$
has the same dimension as the column vector $u$. Therefore
we have $\eta _i\in \R^q$ $(0\le i\le N)$ when $u\in \R ^q$.
Further let $T^r_s(M)$ be the s-times co- and r-times contravariant 
tensor bundle and $(T^r_s)_u(M)$, 
the space of s-times co- and r-times contravariant 
tensors at $u\in M$. We use $X(u)$ (not $X|_u$) to denote a tensor
 of $X\in T_s^r(M)$ at $u\in M$ but sometimes we omit the point $u$ 
for convenience while
there is no confusion of the symbols. 
Note that linear maps $\Phi :T(M)\to T(M)$, $\Psi :T^*(M)\to T^*(M)$, $
J :T^*(M)\to T(M)$, $\Theta :T(M)\to T^*(M)$
may be identified with 
the second degree tensor fields $T_{\Phi}\in T^1_1(M),\ T_{\Psi}\in T^1_1(M),\ 
T_{J}\in T^2_0(M),\ T_{\Theta }\in T^0_2(M)$ 
 by the following relations$^{\cite{Oevel}\,\cite{Manorth}}$
\[\ba {l} T_\Phi (u)(\al (u),K(u))=<\al (u), \Phi (u)K(u)>,\ \al 
\in T^*(M),\ K\in T(M),
\vspace {1mm}\\
 T_\Psi (u)(\al (u),K(u))=<\Psi (u)\al (u), K(u)>,\ \al 
\in T^*(M),\ K\in T(M),
\vspace {1mm}\\
T_J (u)(\al (u),\beta (u))=<\al (u), J (u)\beta (u)>,\ \al ,\beta \in T^*(M),
\vspace {1mm}\\
T_\Theta (u)(K(u),S(u))=<\Theta (u)K(u), S(u)>,\ K,S\in T(M),
 \ea \]
where $<\cdot,\cdot>$ denotes the duality between cotangent vectors and
tangent vectors.

A basic tool to handle various kinds of tensor fields is the 
conception of the Gateaux derivative.
For a tensor fields $X\in T_s^r(M)$, its Gateaux derivative  
at the direction $S\in T(M)$ is defined by
\be X'(u)[S(u)]=\left.\frac {\part X(u+\2 S(u))}{\part \2 }\right|_{\2 =0}.\ee
For four kinds of operators between the tangent bundle and 
the cotangent bundle,
their Gateaux derivatives may be given similarly or 
by means of their tensor fields.
The commutator of two vector fields $K,S\in T(M)$ and the adjoint map
$\textrm{ad}_K$ are defined by
\be [K,S](u)=K'(u)[S(u)]-S'(u)[K(u)],\ \textrm{ad}_KS=[K,S].\ee
The conjugate operator of an operator 
between the tangent bundle and 
the cotangent bundle is established in terms of the duality
between cotangent vectors and
tangent vectors. For example, 
we may calculate the conjugate operator $\Phi ^\dagger :T^*(M)\to T^*(M)$
of an operator $\Phi :T(M)\to T(M)$
through 
\[ <\Phi ^\dagger (u)\al (u) ,K(u) >=<\al (u),\Phi(u) K(u)>,
\ \al \in T^*(M),\ K\in T(M).\]
If an operator $J:T^*(M)\to T(M)$ (or $\Theta :T(M)\to T^*(M)$)
plus its conjugate operator equals to zero, then it is called 
skew-symmetric. 
\begin{definition}
For $H\in C^\infty (M)$, its variational derivative $\frac{\delta H}{\delta u}
\in T^*(M)$
is defined by 
 \[<\frac{\delta H}{\delta u}(u),K(u)>= <\frac{\delta H(u)}{\delta u},K(u)>=H'(u)[K(u)],\ K\in T(M).
\]
If for $\gamma \in T^*(M)$ there exists $H \in  C^\infty (M)$ so that
\[ H'(u)[K(u)]=<\gamma (u)
,K(u)>,\ \textrm{for all}\ K\in T(M),\label{gradient}\]
holds, then $\gamma \in T^*(M)$ is called a gradient field  with 
a potential $H \in  C^\infty (M)$.
\end{definition}
A cotangent vector field $\gamma \in T^*(M)$ is a gradient field  iff 
\be \ba {l}(d\gamma)(u)(K(u),S(u))\vspace{1mm}\\
:= 
  <\gamma '(u)[K(u)],S(u)>-<\gamma '(u)[S(u)],K(u)>=0,\ K,S\in T(M).\ea 
\label{gamma}\ee
If $\gamma \in T^*(M)$ is gradient, then its potential $H\in C^\infty (M)$
is given by 
\[ H(u)=\int _0^1<\gamma (\la u),u>d\lambda.\]
\begin{definition}
A linear operator $\Phi :T(M)\to T(M)$ is called a recursion operator 
of $u_t=K(u),\ K\in T(M),$ if we have for all $S\in T(M)$ and $u\in M$
\be \frac {\part \Phi(u)}{\part t}S(u)+
\Phi '(u)[K(u)]S(u)-K'(u)[\Phi (u)S(u)]+\Phi (u)K'(u)[S(u)]=0.\ee
\end{definition}
Evidently a recursion operator $\Phi :T(M)\to T(M)$ of $u_t=K(u)$ 
maps symmetries into new symmetries of $u_t=K(u)$.
\begin{definition}
A linear operator $\Phi :T(M)\to T(M)$ is called a hereditary 
operator $^{\cite{Fuchssteiner}}$ if the following equality holds
\be\ba {l} \Phi '(u)[\Phi(u) K(u)]S(u)-\Phi(u) \Phi '(u)[K(u)]S(u)
\vspace{1mm}\\ \ 
-\Phi '(u)[\Phi (u)S(u)
]K(u)
+\Phi(u) \Phi '(u)[S(u)]K(u)=0\ea \ee
for all  vector fields $K,S\in T(M)$.
\end{definition}
When an evolution equation $u_t=K(u)$ possesses
a time-independent hereditary recursion operator $\Phi $,
a hierarchy of vector fields
 $\Phi ^nK, \ n\ge 0$, are all symmetries and commute with each other.
If the conjugate operator $\Psi =\Phi ^\dagger $ of 
the hereditary operator $\Phi :T(M)\to T(M)$ maps a gradient field  
$\gamma \in T^*(M)$ into another gradient field , then $\Psi ^n\gamma ,\
 n\ge0$, are all gradient fields$^{\cite{GelfandDorfman}\, \cite{Magri}}$.
\begin{definition} 
A linear skew-symmetric operator
$J:T^*(M)\to T(M)$ is called a Hamiltonian operator
 if for all 
$\alpha , \beta , \gamma \in T^*(M)$ we have
\be <K(u),J'(u)[J(u)S(u)]T(u)>+\textrm{cycle}(K,S,T)=0.\ee The corresponding 
Poisson bracket is defined by 
\be \{H_1,H_2\}_J(u)=<\frac {\delta H_1}{\delta u}(u),J(u)
\frac {\delta H_2}{\delta u}(u)>,\ 
H_1,H_2\in C^\infty(M).\ee
A pair of operators $J,M:T^*(M)\to T(M)$ is called a Hamiltonian
 pair if $J+c M$ is always Hamiltonian
for any constant $c $.
\end{definition}
When $J:T^*(M)\to T(M)$ is Hamiltonian, we 
have$^{\cite{GelfandDorfman}\, \cite{Magri}}$
\[ J\frac {\delta }{\delta u}\{H_1,H_2\}_J=[J\frac {\delta H_1}{\delta u},
J\frac {\delta H_2}{\delta u}],\ H_1,H_2\in C^\infty (M).\]
Moreover if $J,M:T^*(M)\to T(M)$ is a Hamiltonian pair
and $J$ is invertible, then $\Phi =MJ^{-1}:T(M)\to T(M)$ is 
hereditary$^{\cite{FuchssteinerFokas}}$.
\begin{definition}A linear skew-symmetric 
operator $\Theta :T(M)\to T^*(M)$ is called a
symplectic operator if for all $K,S,T\in T(M)$ we have
\be <K(u),\Theta '(u)[S(u)]T(u)>+\textrm{cycle}(K,S,T)=0.\ee
\end{definition}
If $\Theta :T(M)\to T^*(M)$ is 
a symplectic operator, then its second degree tensor field $
T_\Theta \in T_2^0(M)$ may be expressed
as
\[ T_\Theta =d\gamma \ \textrm{with}\  
 <\gamma (u),K(u)>=\int _0^1<\Theta (\la u)\la u, K(u)>\, d\lambda ,
\ K\in T(M),\]
where $d\gamma $ is defined by (\ref{gamma}).
It is not difficult to prove that
 the inverse of symplectic operators are Hamiltonian if they exist
and vice verse. 
\begin{definition}
An evolution equation $u_t=K(u),\ K\in T(M),$ is called a Hamiltonian
equation if there exists a function $H\in C^\infty (M)$ so that 
\be u_t=K(u)=J(u)\frac{\delta H}{\delta u}(u).\ee
 It is called a bi-Hamiltonian 
equation if there exist two functions $H_1,H_2\in C^\infty (M)$ 
and a Hamiltonian pair $J,M:T^*(M)\to T(M)$ so that 
\be u_t=K(u)=J(u)\frac{\delta H_1}{\delta u}(u)=M\frac{\delta H_2}
{\delta u}(u).\ee
\end{definition}
There is another kind of Hamiltonian equations, which 
may be defined by symplectic operators. However, the above definition 
is more advantageous.
For a bi-Hamiltonian equation above, there exist several beautiful  
characteristics in the aspects of algebra and 
geometry$^{\cite{GelfandDorfman}\,\cite{Magri}}$.

In this paper, we shall 
analyze the perturbation equations of the evolution equation 
\begin{equation} u_t=K(u),\ K\in T(M)\label{basicq}\end{equation}
under the perturbation series 
\be \hat{  u}_N=\sum_{i=0}^N \2 ^i\eta _i , \ 
\4 =(\1 _0^T,\1 _1^T.\cdots,\1 _N^T)^T,\ 
N\ge 1
\label{pseries}\ee 
and their integrable properties.
The paper is organized as follows.
In Section 2 we propose three useful theorems to generate Hereditary 
operators, Hamiltonian operators and symplectic operators
 from the perturbation series (\ref{pseries})
in a natural and explicit way.
We show in Section 3 that the perturbation (\ref{pseries})
preserves complete integrability,
i.e. we want to show that the perturbation equations are still
integrable equations provided that the equation under consideration
is integrable. In Section 4, we apply the resulting theory to
 MKdV hierarchy and KP equation as illustrative examples.
Section 5 contains some concluding remarks, where we give another 
perturbation series and compare its corresponding results with
ones by (\ref{pseries}).

\section{Hereditary, Hamiltonian or symplectic operators by perturbation}
\setcounter{equation}{0}

We make a perturbation series for any $m\ge 0$
\be
\hat{  u}_m=\sum_{i=0}^m \2 ^i\eta _i , \ 
\hat {\1 }_m =(\1 _0^T,\1 _1^T,\cdots,\1 _m^T)^T,\ee
where $\eta _i,\ 0\le i\le m,$ are all column vectors possessing the 
same dimension as $u$. We first analyze a few of properties of tensor 
fields and then 
establish three  useful theorems to construct 
hereditary, Hamiltonian or symplectic operators in terms of a perturbation.

\begin{lemma}\label{same} 
We have for any $X\in T_s^r(M)$
\begin{equation} 
\frac {\part ^k X(\hat {u}_i)}{\part \2 ^k}=
\frac {\part ^k X(\hat {u}_j)}{\part \2 ^k},\ i,j\ge k\ge 0.
\end{equation}
\end{lemma}
{\bf Proof:} Let  $j>i$ without loss of generality. Then we have
\[ \hat {u}_j=\hat {u}_i +\sum_{l=i+1}^j\2 ^l\1 _l.\]
Further we get
\[ X(\hat{u}_j)=X(\hat{u}_i)+\textrm{o}(\2 ^i),\]
from which the required equality follows. \#

Let us assume for any $X\in T^r_s(M)$ 
\begin{equation}
\bigr(X(u)\bigl)^{(i)}(\hat {\1 }_j)= \frac1 {i!}
\left.\frac {\part ^i X(\hat {u}_j)}{\part \2 ^i}\right | _{\2 =0}, \ 
X^{(i)}=(X(u))^{(i)}=\bigr(X(u)\bigl)^{(i)}(\hat {\1 }_i),\ j\ge i\ge0.
\end{equation}
We write \be (\textrm{per}_mX)(\hat {\eta }_m)=
\hat {X}_m(\hat {\eta }_m)=(X^{(0)T}(\hat {\eta }_m),X^{(1)T}(\hat {\eta }_m)
,\cdots,X^{(m)T}(\hat {\eta }_m))^T\ \,( m\ge0)\ee
 and 
call $\textrm{per}_mX=\hat{X}_m$
 the perturbation tensor field of order $m$. From Lemma \ref{same},
we see that 
\[\hat {X}_m(\hat{\eta}_m) =(X^{(0)T}(\hat{\eta}_0),X^{(1)T}
(\hat{\eta}_1),\cdots,X^{(m)T}(\hat{\eta}_m))^T\]
 and
 hence the perturbation tensor field $\textrm{per}_mX=\hat{X}_m$ 
has a characteristic:
the $i$-th component depends only on $\eta _0,\eta _1,\cdots , \eta _i$,
not on any $\eta _j,\ j>i$.

\begin{lemma} Let $X\in T_s^r(M)$ and 
$S\in T(M)$, we have
\begin{equation} 
\bigr(X'(u)[S(u)]\bigl)^{(i)}(\hat {\1 }_i)=\Bigr(\bigr(X(u)\bigl)^{(i)}
\Bigl)'(\hat {\1 }_i)[\hat {S }_i],\ i\ge0,\label {lemma2eq1}
\end{equation}
where
$\hat {S }_i=(S^{(0)T},S^{(1)T},\cdots,S^{(i)T})^T$.
\end{lemma}
{\bf Proof:} We first have 
\begin{eqnarray} 
&&(X(\hat {u}_i))'(\hat {\eta}_i)[\hat{S}_i]=\left.
\frac {\part }{\part \delta}\right|_{\delta =0}X(\hat {u}_i
+\delta \sum_{k=0}^i\2 ^kS^{(k)})
\nonumber\\
&=& \left.\frac {\part }{\part \delta}\right|_{\delta =0}K(\hat {u}_i+\delta 
S(\hat {u}_i)+\delta \textrm{o}(\2 ^i))\nonumber\\
&=& X'(\hat {u}_i)[S(\hat {u}_i)] +\textrm{o}(\2 ^i).\nonumber
\end{eqnarray}
We apply the above equality to the following Taylor series 
\[
X(\hat {u}_i)=\sum_{k=0}^i\frac{\2 ^k}{i!}
\left.\frac{\part ^k X(\hat {u}_i)}{\part \2 ^k}
\right  |_{\2 =0}+\textrm{o}(\2 ^i)\]
and then get the required equality (\ref{lemma2eq1}).
\#

Evidently, (\ref{lemma2eq1}) implies that for $ X\in T^r_s(M),\ S\in T(M)$ we 
have
\[(\textrm{per}_m
(X'(u)[S(u)]))(\hat {\1 }_m)= 
(\textrm{per}_mX(\hat{\1 }_m))'(\hat{\1 }_m)[(\textrm{per}_mS)(\hat {\1 }
_m)].\]

\begin{lemma}  Let $X\in T_s^r(M)$.
 The following equalities hold for any vector 
field $\bar {S}_N=(S_0^T,S_1^T,\cdots,S_N^T)^T\in T(\hat{M}_N)$,
where $S_i,\ 0\le i\le N$, are of the same dimension,
\begin{equation}
\Bigl (\left.\frac {\part ^iX (\hat{u}_N)}{\part \2 ^i}\right|_{\2 =0}
\Bigr )'(\hat {\1 }_N)[\bar S_N]=\left.
\frac{\part ^i}{\part \2 ^i}\right|_{\2  =0
}X'(\hat {u}_N)\bigl[\sum_{
j=0}^N\2 ^jS_j\bigr],\ 0\le i\le N.\label{lemma3eq1}
\end{equation}
\end{lemma}
{\bf Proof:}
First from Taylor series \[
X(\hat {u}_N)=\sum_{i=0}^N\frac {\2 ^i}{i!}
\left.\frac{\part ^i X(\hat {u}_N)}{\part \2 ^i}
\right  |_{\2 =0}+\textrm{o}(\2 ^N),\]
we directly obtain
\[(X(\hat{u}_N))'(\4 )[\bar{S}_N]=\sum_{i=0}^N\frac {\2 ^i}{i!}
\Bigl(\left.\frac {\part ^iX(\hat{u}_N)}{\part \2 
^i}\right|_{\2 =0}\Bigr)'(\4 )[
\bar{S}_N]+\textrm{o}(\2 ^N).\]
On the other hand, we have 
\[(X(\hat{u}_N))'(\4 )[\bar{S}_N]=
\left.\frac {\part }{\part \delta}\right|_{\delta =0}X(\hat {u}_N+\delta
\sum_{j=0}^N\2 ^jS_j)=
X'(\hat {u}_N )\Bigl [ \sum_{j=0}^N\2 ^j S_j\Bigr].\]
These two equalities give (\ref{lemma3eq1}) again according to Taylor series.
The proof is finished. \#

\begin{theorem}
If the operator $\Phi :T(M)\to T(M)$ is  hereditary, 
then the  operator $\P : T(\hat{M}_N)\to T(\hat{M}_N)$ defined by 
\begin{eqnarray}
&&(\textrm{per}_N\Phi)(\hat{\1 }_N)= \P (\4 )
\nonumber
\\&=&\left[ \bigl(\P (\4 )\bigr)_{ij}\right]_{i,j=0,1,\cdots,N}
=\left[
\frac 1 {(i-j)!}\left.\frac {\part ^{i-j}\Phi (\3 )}{\part \2 ^{i-j}}
\right|_{\2 =0}
 \right]_{(N+1)\times(N+1)}\nonumber
\\
 &= &
\left[ \begin{array}{cccc}
\Phi (\1 _0)& & &0\vspace{1mm}\\ 
\frac {1}{1!}\left.\frac {\part \Phi (\3 )}
{\part \2 }\right| _{\2 =0}&\Phi(\1 _0
) & 
&\vspace{1mm} \\
\vdots & \ddots & \ddots &\vspace{1mm} \\
\frac {1}{N!}\left.\frac {\part ^N \Phi (\3 )}{\part \2 ^N }
\right|_{\2 =0}&\cdots &
\frac {1}{1!}\left.\frac {\part \Phi (\3 )}{\part \2 }
\right|_{\2 =0}&\Phi(\1 _0)
\end{array}
\right]\label{newhere}
\end{eqnarray}
is hereditary, too.\label{hereditary}
\end{theorem}
{\bf Proof:} Let $\bar{ K}_N=(K_0^T,K_1^T,\cdots,K_N^T)^T,$ $
\bar{ S}_N=(S_0^T,S_1^T,\cdots,S_N^T)^T\in T(\hat{M}_N)$,
where $K_i,S_i,\ 0\le i\le N$, are of the same dimension.
It suffices to prove that
\begin{eqnarray}
&&\P '(\4 )[\P \bar{ K}_N]\bar{ S}_N-\P \P '(\4 )[\bar{ K}_N]
\bar{ S}_N\nonumber\\
 &&-
\P '(\4 )[\P \bar{ S}_N]\bar{ K}_N+\P \P '(\4 )[\bar{ S}_N]\bar{ K}_N=0.
\label{heequiv}
\end{eqnarray}
First we easily get the $i$-th element of the vector field
$\P \bar{K}_N $ and the element in the $(i,j)$ position of the matrix
$\P '(\4 )[\bar{K}_N]$:
 \begin{eqnarray}
(\P 
\bar{K}_N
)_i=\sum_{j=0}^i
\frac1 {(i-j)!}\left.\frac {\part ^{i-j}\Phi (\3 )}{\part \2 ^{i-j}}
 \right|_{\2 =0} K_j,\ 0\le i\le N,&&\nonumber\\
\bigl(\P '(\4 )[\bar{K}_N]\bigr)_{ij}=
\frac1 {(i-j)!}\left(\left.
\frac {\part ^{i-j}\Phi (\hat {u}_N)}{\part \2 ^{i-j}}
 \right|_{\2 =0}\right)'(\4 )[\bar{K}_N]&&\nonumber\\
\qquad  =
\frac1 {(i-j)!}\left.\frac {\part ^{i-j}}{\part \2 ^{i-j}}
 \right|_{\2 =0}\Bigl(\Phi '(\3 )\bigl[\sum_{k=0}^N\2 ^kK_k\bigr]\Bigr)
 ,\ 0\le i,j\le N.&&\nonumber\end{eqnarray}
Here we use Lemma 2.3 for the calculation of the second equality.
Now  we can compute the $i$-th element of $\P '(\4 )[\P \bar{K}_N]\bar{S}_N
$:
\begin{eqnarray}
& &(\P '(\4 )[\P \bar{K}_N]\bar{S}_N)_i\nonumber\\&=&
\sum_{j=0}^i
\frac1 {(i-j)!}\left.\frac {\part ^{i-j}}{\part \2 ^{i-j}}
 \right|_{\2 =0}
\Phi '(\3 )\Bigl[
\sum_{k=0}^N\2 ^k \sum_{l=0}^k 
\frac1 {(k-l)!}\left.\frac {\part ^{k-l}\Phi (\3 )}{\part \2 ^{k-l}}
 \right|_{\2 =0}K_l
\Bigr]S_j\nonumber\\
& = &
\sum_{j=0}^i
\frac1 {(i-j)!}\left.\frac {\part ^{i-j}}{\part \2 ^{i-j}}
 \right|_{\2 =0}
\Phi '(\3 )\Bigl[
\sum_{l=0}^N\2 ^l \sum_{k=l}^N 
\frac{\2 ^{k-l}} {(k-l)!}\left.\frac {\part ^{k-l}\Phi (\3 )}{\part \2 ^{k-l}}
 \right|_{\2 =0}K_l
\Bigr]S_j\nonumber\\
& =&
\sum_{j=0}^i
\frac1 {(i-j)!}\left.\frac {\part ^{i-j}}{\part \2 ^{i-j}}
 \right|_{\2 =0}
\Phi '(\3 )\Bigl[
\sum_{l=0}^N\2 ^l \bigl(\Phi (\3 )+\textrm{o}(\2 ^{N-l})\bigr)K_l
\Bigr]S_j\nonumber\\
& =&
\sum_{j=0}^i
\frac1 {(i-j)!}\left.\frac {\part ^{i-j}}{\part \2 ^{i-j}}
 \right|_{\2 =0}
\Phi '(\3 )\Bigl[
\sum_{l=0}^N\2 ^l \bigl(\Phi (\3 )\bigr)K_l
\Bigr]S_j\nonumber\\
&=&\sum_{0\le j+l\le i}
\frac1 {(i-j-l)!}\left.\frac {\part ^{i-j-l}}{\part \2 ^{i-j-l}}
 \right|_{\2 =0}
\Phi '(\3 )\Bigl[\Phi (\3 )K_l\Bigr]S_j,\  0\le i\le N,\nonumber
\end{eqnarray}
and the $i$-th element of $\P \P '(\4 )[\bar {K}_N]\bar {S}_N$:
\begin{eqnarray}
&&\bigl(\P \P '(\4 )[\bar {K}_N]\bar {S}_N\bigr)_i\nonumber
\\&=&
\sum_{j=0}^i\frac1 {(i-j)!}\left.
\frac {\part ^{i-j}\Phi (\3 )}{\part \2 ^{i-j}}
 \right|_{\2 =0}\sum_{k=0}^j
\frac1 {(j-k)!}\left.\frac {\part ^{j-k}}{\part \2 ^{j-k}}
 \right|_{\2 =0}\Phi '(\3 )\Bigl[\sum_{l=0}^N\2 ^lK_l\Bigr]S_k\nonumber\\
& =&\sum_{j=0}^i\frac1 {(i-j)!}\left.
\frac {\part ^{i-j}\Phi (\3 )}{\part \2 ^{i-j}}
 \right|_{\2 =0}
\sum_{ k=0}^j\sum_{l=0}^{j-k}
\frac1 {(j-k-l)!}\left.\frac {\part ^{j-k-l}}{\part \2 ^{j-k-l}}
 \right|_{\2 =0}\Phi '(\3 )[K_l]S_k\nonumber\\
& =&\sum_{k=0}^i\sum_{j=k}^i\sum_{l=0}^{j-k}
\frac1 {(i-j)!(j-k-l)!}\left.\frac {\part ^{i-j}\Phi (\3 )}{\part \2 ^{i-j}}
 \right|_{\2 =0}\left.\frac {\part ^{j-k-l}}{\part \2 ^{j-k-l}}
 \right|_{\2 =0}\Phi '(\3 )[K_l]S_k\nonumber\\
& =&\sum_{k=0}^i\sum_{l=0}^{i-k}\sum_{j=k+l}^i
\frac1 {(i-j)!(j-k-l)!}\left.\frac {\part ^{i-j}\Phi (\3 )}{\part \2 ^{i-j}}
 \right|_{\2 =0}\left.\frac {\part ^{j-k-l}}{\part \2 ^{j-k-l}}
 \right|_{\2 =0}\Phi '(\3 )[K_l]S_k\nonumber\\
& =&\sum_{k=0}^i\sum_{l=0}^{i-k} 
\frac1 {(i-k-l)!}\left.\frac {\part ^{i-k-l}}{\part \2 ^{i-k-l}}
  \right|_{\2 =0}\Bigl(\Phi (\3 )\Phi '(\3 ) [K_l]S_k\Bigr)\nonumber\\
& =&\sum_{0\le k+l\le i}
\frac1 {(i-k-l)!}\left.\frac {\part ^{i-k-l}}{\part \2 ^{i-k-l}}
  \right|_{\2 =0}\Bigl(\Phi (\3 )\Phi '(\3 ) [K_l]S_k\Bigr),\ 0\le i\le N
.\nonumber
\end{eqnarray}
Therefore by the hereditary property of $\Phi$,
we find that each element in the left side of 
(\ref{heequiv}) is zero, which shows (\ref{heequiv}) holds.
The proof is completed. \#

\begin{theorem}
If the operator $J :T^*(M)\to T(M)$ is Hamiltonian,
 then the operator $\J :T^*(\hat{M}_N)\to T(\hat{M}_N)$ defined by
\begin{eqnarray}&&(\textrm{per}_NJ)(\hat{\1 }_N)=
\J (\4 )\nonumber
\\
&=&\left[ \bigl(\J (\4 )\bigr)_{ij}\right]_{i,j=0,1,\cdots,N}
=\left[
\frac 1 {(i+j-N)!}\left.\frac {\part ^{i+j-N}J (\3 )}{\part \2 ^{i+j-N}}
\right|_{\2 =0}
 \right]_{(N+1)\times(N+1)}\nonumber
\\
 &= &
\left[ \begin{array}{cccc}
0& & & J (\1 _0)\\ & &\vdots& 
\frac 1{1!}\left.
\frac {\part  J (\3 )}{\part \2 }
\right|_{\2 =0} \\
 &\vdots &\vdots& \vdots\\ J (\1 _0)&\frac 1{1!}\left.
\frac {\part  J (\3 )}{\part \2 }
\right|_{\2 =0}&\cdots&
\frac {1}{N!}\left.\frac {\part ^N J (\3 )}{\part \2 ^N }
\right|_{\2 =0} 
\end{array}
\right]\label{newham}
\end{eqnarray}is Hamiltonian, too.
\end{theorem}
{\bf Proof:} 
Let $\bar{\al }_N=(\al _0^T,\al _1^T,\cdots,\al _N^T)^T$, $\bar{\beta
 }_N=(\beta  _0^T,\beta  _1^T,
\cdots,\beta  _N^T)^T$,
$\bar{\gamma  }_N=(\gamma  _0^T,\gamma  _1^T,\cdots,\gamma _N^T)^T\in T^*(\hat{M}_N)$,
where $\al _i,\beta _i,\gamma _i\ 0\le i\le N$, are of the same dimension.
We only need to prove that
\begin{equation}
<\bar{\al }_N,
\hat {J}_N 
'(\4 )[\hat {J}_N \bar{\beta  }_N]\bar{ \gamma }_N>+\textrm{cycle}
(\bar{\al }_N,\bar{\beta  }_N,
\bar{\gamma  }_N)=0,
\end{equation}
because there doesn't exist any problem on linearity and
the skew-symmetric property.
First noting Lemma 2.3,  we can compute the element in the $(i,j)$ 
position of the matrix $
\hat {J}_N '(\4 )[\hat {J}_N \bar{\beta  }_N]$:
\begin{eqnarray} 
&&\bigl(\hat {J}_N '(\4 )[\hat {J}_N \bar{\beta  }_N]\bigr)_{ij}\nonumber\\
& =& \frac 1 {(i+j-N)!}\left.\frac {\part ^{i+j-N}}{\part \2 ^{i+j-N}}
\right|_{\2 =0} {J} '(\3 )\Bigl[ \sum _{l=0}^N\2 ^l 
(\hat {J}_N \bar{\beta  }_N)_l\Bigr]\nonumber\\
&= &\frac 1 {(i+j-N)!}\left.\frac {\part ^{i+j-N}}{\part \2 ^{i+j-N}}
\right|_{\2 =0} {J} '(\3 )\Bigl[ \sum _{l=0}^N\2 ^l 
 \sum_{k=N-l}^N
\frac 1 {(k+l-N)!}\left.\frac {\part ^{k+l-N}
 {J}(\3 )}{\part \2 ^{k+l-N}}
\right|_{\2 =0}\beta _k
\Bigr]\nonumber\\
&=&\frac 1 {(i+j-N)!}\left.\frac {\part ^{i+j-N}}{\part \2 ^{i+j-N}}
\right|_{\2 =0}J '(\3 )\Bigl[\sum_{k=0}^N\2 ^{N-k}\bigl(\sum_{l=N-k}
^{N}\frac {\2 ^{k+l-N}}{(k+l-N)!}\left.\frac{\part ^{k+l-N}J(\3 )}
{\part \2 ^{k+l-N}}
\right|_{\2 =0}\bigr)\beta _k\Bigr]\nonumber\\
&=&\frac 1 {(i+j-N)!}\left.\frac {\part ^{i+j-N}}{\part \2 ^{i+j-N}}
\right|_{\2 =0}{J}'(\3 )\Bigl[\sum_{k=0}^N\2 ^{N-k}\bigl(
J(\hat{u}_N )+\textrm{o}(\2 ^k)\bigr)\beta _k\Bigr]\nonumber\\
&=&\frac 1 {(i+j-N)!}\left.\frac {\part ^{i+j-N}}{\part \2 ^{i+j-N}}
\right|_{\2 =0} {J} '(\3 )\Bigl[\sum_{k=0}^N\2 ^{N-k}
J(\3 )\beta _k\Bigr]\nonumber\\
&=&\sum_{k=0}^N\frac 1 {(i+j-N)!}\left.\frac {\part ^{i+j-N}}
{\part \2 ^{i+j-N}}
\right|_{\2 =0}\bigl(\2 ^{N-k} {J} '(\3 )\bigl[
J(\3 )\beta _k\bigr]\bigr)\nonumber\\
&=&\sum_{k=2N-(i+j)}^N\frac 1 {(i+j+k-2N)!}\left.
\frac {\part ^{i+j+k-2N}}{\part \2 ^{i+j+k-2N}}
\right|_{\2 =0}\bigl( {J} '(\3 )\bigl[
J(\3 )\beta _k\bigr]\bigr),\ 0\le i,j\le N.\nonumber
\end{eqnarray}
Therefore we have 
\begin{eqnarray} &&<\bar{\al }_N,
\hat {J}_N 
'(\4 )[\hat {J}_N \bar{\beta  }_N]\bar{ \gamma  }_N>+\textrm{cycle}(\bar{\al }_N,\bar{\beta  }_N,
\bar{\gamma  }_N)\nonumber\\
&= &\sum_{2N\le i+j+k\le 3N}\frac 1 {(i+j+k-2N)!}\left.
\frac {\part ^{i+j+k-2N}}{\part \2 ^{i+j+k-2N}}
\right|_{\2 =0}\Bigl(<\al _i, {J} '(\3 )\bigl[
J(\3 )\beta _k\bigr]\gamma _j>\nonumber\\
& & \qquad +\textrm{cycle}(\al _i,\beta _k,\gamma _j)\Bigr)=0.
\nonumber
\end{eqnarray}
In the last step, we have utilized the Hamiltonian property of $J(u)$. 
So the required result is proved. \#

Completely similar to the above two theorems, we can show the following result.

\begin{theorem}
If the operator $\Theta :T(M)\to T^*(M)$ is symplectic,
 then the operator $\hat{\Theta }_N :T(\hat{M}_N)\to T^*(\hat{M}_N)$ defined by
\begin{eqnarray}&&(\textrm{per}_N\Theta )(\hat{\1 }_N)=
\hat{\Theta }_N (\4 )\nonumber
\\
&=&\left[ \bigl(\hat{\Theta }_N (\4 )\bigr)_{ij}\right]_{i,j=0,1,\cdots,N}
=\left[
\frac 1 {(N-i-j)!}\left.\frac {\part ^{N-i-j}\Theta (\3 )}{\part \2 ^{N-i-j}}
\right|_{\2 =0}
 \right]_{(N+1)\times(N+1)}\nonumber
\\
 &= &
\left[ \begin{array}{cccc}
\frac {1}{N!}\left.\frac {\part ^N \Theta (\3 )}{\part \2 ^N }
\right|_{\2 =0}
&\cdots &\frac 1{1!}\left.
\frac {\part  \Theta (\3 )}{\part \2 }
\right|_{\2 =0} 
 & \Theta (\1 _0)\vspace{1mm}\\  \vdots &\vdots &\vdots& \vspace{1mm} \\
\frac {1}{1!}\left.\frac {\part ^N \Theta (\3 )}{\part \2 ^N }
\right|_{\2 =0} 
 &\vdots & &  \vspace{2mm}\\ \Theta (\1 _0)& & & 0
\end{array}
\right]\label{newsym}
\end{eqnarray}is symplectic, too.
\end{theorem}

We mention that when
$J$ and $\Theta $ are invertible and $J=\Theta ^{-1}$, 
we have 
\[ (\textrm{per}_NJ)(\hat{\eta }_N)=((\textrm{per}_N\Theta )(\hat{\eta }_N))^{-1},
\]
which shows that the inverse of the perturbation symplectic
operator $(\textrm{per}_N\Theta )(\hat{\eta }_N)$ is Hamiltonian 
and vice versa. Note that the above 
new operators have a vector $\hat {\eta }_N$ of dependent variables
and hence involve $q(N+1)$ dependent variables when $u$ is a $q$-dimensional
vector.
The above three theorems also provide us a method to generate new 
Hereditary, Hamiltonian or symplectic 
operators from a known one. This is interesting
in the soliton theory. In particular, we can 
put forward the following operators
by the first order perturbation:
\begin{eqnarray}&&
(\hat{\Phi}_1) (\hat {\eta }_1)=\left[
\ba {cc} \Phi (\eta _0)&0\\ \Phi '(\eta _0)[\eta _1]&
\Phi (\eta _0)\ea \right],
\nonumber\\&&
 (\hat{J}_1) (\hat {\eta }_1)=\left[
\ba {cc}0& J (\eta _0)\\ J (\eta _0)&J '(\eta _0)[\eta _1]
\ea \right],
\nonumber\\ &&
(\hat{\Theta }_1) (\hat {\eta }_1)=\left[
\ba {cc}\Theta '(\eta _0)[\eta _1]& \Theta (\eta _0)\\ \Theta (\eta _0)& 0
\ea \right]. \nonumber
\end{eqnarray}
Taking the first order perturbation once more, we can obtain a little 
more complicated
operators:
\begin{eqnarray}&&
(\textrm{per}_1\textrm{per}_1\Phi) (\hat {\eta }_3)
=\left[\ba {cccc} \Phi (\eta _0)&0&0&0\vspace{1mm}\\
\Phi '(\eta _0)[\eta _1]& \Phi (\eta _0)&0&0\vspace{1mm}\\
\Phi '(\eta _0)[\eta _2]&0&\Phi '(\eta _0)[\eta _1]&\Phi (\eta _0)
\vspace{1mm}\\
\left.
\frac{\part ^2\Phi (\eta _0+\varepsilon \eta _1+\delta \eta _2+\varepsilon
\delta \eta _3)}{\part \varepsilon \part \delta }\right|_{\varepsilon=\delta=0}
&\Phi '(\eta _0)[\eta _2]&\Phi '(\eta _0)[\eta _1]& \Phi (\eta _0)
\ea \right],
\nonumber\\ &&
(\textrm{per}_1\textrm{per}_1J) (\hat {\eta }_3)
=\left[\ba {cccc}0&0&0& J (\eta _0)\vspace{1mm}\\ 0&0&J (\eta _0)&
J '(\eta _0)[\eta _1]\vspace{1mm}\\0&J (\eta _0)&0&
J '(\eta _0)[\eta _2]\vspace{1mm}\\
J (\eta _0)&J '(\eta _0)[\eta _1]&J '(\eta _0)[\eta _2]
&
\left.
\frac{\part ^2J (\eta _0+\varepsilon \eta _1+\delta \eta _2+\varepsilon
\delta \eta _3)}{\part \varepsilon \part \delta }\right|_{\varepsilon=\delta=0}
\ea \right],
\nonumber\\ 
&&(\textrm{per}_1\textrm{per}_1\Theta ) (\hat {\eta }_3)=\left[
\ba {cccc}
\left.
\frac{\part ^2\Theta (\eta _0+\varepsilon \eta _1+\delta \eta _2+\varepsilon
\delta \eta _3)}{\part \varepsilon \part \delta }\right|_{\varepsilon=\delta=0}
& \Theta '(\eta _0)[\eta _2]&\Theta '(\eta _0)[\eta _1]&\Theta (\eta _0)
\vspace{1mm}\\
\Theta '(\eta _0)[\eta _2]&0&\Theta (\eta _0)&0\vspace{1mm}\\
\Theta '(\eta _0)[\eta _1]&\Theta (\eta _0)&0&0\vspace{1mm}\\
\Theta (\eta _0)&0&0&0\ea \right]
\nonumber.
\end{eqnarray}
Here we have changed two dependent variables while making the second 
perturbation. Of course this kind of perturbation may be done without 
any stop at finite steps and hence the resulting operators are 
full of various algebraic structures.

\section{Integrable properties of the perturbation equations}
\label{integrableproperties}
\setcounter{equation}{0}

Let us recall the perturbation series (\ref{pseries})
\[
\hat{  u}_N=\sum_{i=0}^N \2 ^i\eta _i , \ 
\4 =(\1 _0^T,\1 _1^T.\cdots,\1 _N^T)^T,\ 
N\ge 1.\]
For the evolution equation (\ref{basicq})
\[
u_{t}=K(u),\ K\in T(M),\]
we consider  the following $N$-th order perturbation equation 
 \begin{equation}
\hat {u}_{Nt}=K(\3 )+\textrm{o}(\2 ^N)\ \ \textrm{or}\ \ 
\hat {u}_{Nt}=K(\3 )\quad \pmod{\2 ^N},\end{equation}
which leads to the following equivalent equation
\begin{equation}
 \hat {\1 }_{Nt}=\hat {K}_{N}(\4 ),\ \  \textrm{namely}\ \ 
\1 _{it}=\left.
\frac1 {i!} \frac {\part K(\3 )}{\part \2 ^i}\right|_{\2 =0},\ 0\le i\le N.
\label{pe}
\end{equation}
  In this section, we would like to discuss integrable properties of 
the perturbation equation (\ref{pe}), which include recursion operators,
$K$-symmetries (i.e. time independent symmetries), master symmetries,
linear representations (Lax representation
and zero curvature representation) and 
Hamiltonian structures etc.
\begin{theorem} Let $K\in T(M)$.
The operator $\P : T(\hat {M}_N)\to T(\hat {M}_N)$ determined by 
(\ref{newhere})
 is a recursion operator of the perturbation equation
$\hat {\1 }_{Nt}=
\5 (\4 )$ defined by (\ref{pe}) when $\Phi : T(M)\to T(M)$
 is a recursion operator of $u_t=K(u)$.
Therefore the perturbation equation $\hat {\1 }_{Nt}=
\5 (\4 )$ has a hereditary recursion operator 
$\P (\4 )$ once $u_t=K(u)$  has a hereditary recursion operator $\Phi (u)$.
\end{theorem}
{\bf Proof:} Let $\bar{S}_N=(S_0^T,S_1^T,\cdots,S_N^T)^T\in T(\hat {M}_N)$,
where $S_i,\ 0\le i\le N$, are of the same dimension.
 By Lemma 2.3, we have
\begin{eqnarray}
&& 
\left(\left. \frac {\part ^k\Phi (\3 )}{\part \varepsilon ^k}
\right|_{ \varepsilon=0}\right)'(\4 )[\5 ]=\left.\frac {\part ^k}{\part 
\varepsilon ^k}\right|_{\2 =0}
\Phi '(\hat {u}_N)\Bigl[\sum_{j=0}^N\2 ^j K^{(j)}\Bigr]
\nonumber\\
 &=& \left.\frac {\part ^k}{\part 
\varepsilon ^k}\right|_{\2 =0}
\Phi '(\hat {u}_N)\bigl[K(\hat {u}_N)+\textrm{o}(\2 ^N)\bigr]=\left.
\frac {\part ^k \Phi '(\hat {u}_N)[K(\hat {u}_N)]}{\part \2 ^k}\right|_{\2 =0}
,\ 0\le k\le N
,\nonumber
\end{eqnarray}
and 
\begin{eqnarray}
&&(K^{(i)})'(\4 )[\bar{S}_N]
=\frac 1 {i!}\left.\frac {\part ^i}{\part \2 ^i}
\right|_{\2 =0}K'(\3 )\Bigl[\sum _{k=0}^N\2 ^kS_k\Bigr]\nonumber\\
&=& \sum_{j=0}^i\frac1 {(i-j)!}
\left.\frac {\part ^{i-j} K'(\3 )[S_j]}{\part \2 ^{i-j}}
\right|_{\2 =0},\ 0\le i\le N.\nonumber
\end{eqnarray}
Therefore from the above first equality, we obtain
 the $i$-th element of $\P ' (\4 )[\5 ]\bar{S}_N$: 
\begin{equation}
\bigl( \P ' (\4 )[\5 ]\bar{S}_N \bigr)_i=\sum_{j=0}^i\frac1 {(i-j)!}\left.
\frac{\part ^{i-j} \Phi '(\3 )[K(\3 )]S_j}{\part \2 ^{i-j}}\right|_{\2 =0},\ 
0\le i\le N.\label{recursion1}
\end{equation}
From the above second equality, we can compute the following two terms:
\begin{eqnarray}
&&\bigl(\5 '(\4 )[\P \bar {S}_N]\bigr)_i\nonumber\\
&=&\sum_{k=0}^i\frac1 {(i-k)!}
\left.
\frac{\part ^{i-k} 
 }{\part \2 ^{i-k}}\right|_{\2 =0}\bigl( K
 '(\3 )[(\P \bar{S}_N)_k]\bigr)\nonumber\\
&=&\sum_{k=0}^i\frac1 {(i-k)!}
\left.
\frac{\part ^{i-k} 
 }{\part \2 ^{i-k}}\right|_{\2 =0}K'(\3 )\Bigl[\sum_{j=0}^k\frac1 {(k-j)!}
\left.
\frac{\part ^{k-j} \Phi (\3 )
 }{\part \2 ^{k-j}}\right|_{\2 =0}S_j\Bigr]\nonumber\\
&=& 
\sum_{j=0}^i\sum_{k=j}^i\frac 1{(i-k)!(k-j)!}\left.\frac {\part ^{i-k}}{
\part \2 ^{i-k}}\right|_{\2 =0}K'(\3 )\Bigl[
\left.
\frac{\part ^{k-j} \Phi (\3 )
 }{\part \2 ^{k-j}}\right|_{\2 =0}S_j\Bigr]\nonumber\\
&=& 
\sum_{j=0}^i\sum_{k=j}^i\frac 1{(i-j)!}\left.\frac {\part ^{i-j}}{
\part \2 ^{i-j}}\right|_{\2 =0}\Bigl(\frac {\2 ^{k-j}}{(k-j)!}
K'(\3 )\Bigl[
\left.
\frac{\part ^{k-j} \Phi (\3 )
 }{\part \2 ^{k-j}}\right|_{\2 =0}S_j\Bigr]\Bigr)\nonumber\\
&=& 
\sum_{j=0}^i\frac 1{(i-j)!}\left.\frac {\part ^{i-j}}{
\part \2 ^{i-j}}\right|_{\2 =0}
K'(\3 )\Bigl[\sum_{k=j}^i\frac {\2 ^{k-j}}{(k-j)!}
\left.
\frac{\part ^{k-j} \Phi (\3 )
 }{\part \2 ^{k-j}}\right|_{\2 =0}S_j\Bigr]\nonumber\\
 &=&\sum_{j=0}^i\frac1 {(i-j)!}
\left.
\frac{\part ^{i-j} 
 }{\part \2 ^{i-j}}\right|_{\2 =0}K'(\3 )[
\Phi (\3 )S_j+\textrm{o}(\2 ^{i-j})]\nonumber\\
&=&\sum_{j=0}^i\frac1 {(i-j)!}
\left.
\frac{\part ^{i-j} \bigl(K'(\3 )[\Phi (\3 )S_j]\bigr)
 }{\part \2 ^{i-j}}\right|_{\2 =0}, \ 0\le i\le N;\label{recursion2}
\end{eqnarray}
\begin{eqnarray}
&& \bigl(\P \5 '(\4 )[\bar{S}_N]\bigr)_i\nonumber\\
&=&\sum_{k=0}^i\frac1 {(i-k)!}
\left.
\frac{\part ^{i-k} \Phi (\3 )
 }{\part \2 ^{i-k}}\right|_{\2 =0}
\sum_{j=0}^k\frac1 {(k-j)!}
\left.
\frac{\part ^{k-j} K'(\3 )[S_j]
 }{\part \2 ^{k-j}}\right|_{\2 =0}
\nonumber\\
&=&\sum_{j=0}^i\sum_{k=j}^i\frac1 {(i-k)!(k-j)!}
\left.
\frac{\part ^{i-k} \Phi (\3 )
 }{\part \2 ^{i-k}}\right|_{\2 =0}
\left.
\frac{\part ^{k-j} K'(\3 )[S_j]
 }{\part \2 ^{k-j}}\right|_{\2 =0}\nonumber\\
&=&\sum_{j=0}^i\frac1 {(i-j)!}
\left.
\frac{\part ^{i-j} \bigl(\Phi (\3 )K'(\3 )[S_j]\bigr)
 }{\part \2 ^{i-k}}\right|_{\2 =0},\ 0\le i\le N.\label {recursion3}
\end{eqnarray}
Now in virtue of 
 the above three equalities (\ref{recursion1}), (\ref{recursion2}) and
(\ref{recursion3}), we easily see that
\[\frac {\part \hat {\Phi}_N}{\part t}\bar {S}_N+
 \P '(\4 )[\5 ]\bar{S}_N-\5 '(\4 )[\P \bar{S}_N]+\P \5 '(\4 ) 
[\bar{S}_N]=0,\]
which implies that $\hat{\Phi}_N (\hat {\eta }_N)$ is a recursion operator
of $\hat {\eta }_{Nt}=\hat {K}_N$. A combination 
with Theorem \ref{hereditary} gives the proof of the second conclusion.
\#

\begin{theorem} Let $K,S\in T(M)$.
There exists  a relation between the perturbation vector fields
 $\5 $ and $\6 $
\begin{equation}
[\5 (\4 ), \6 (\4 )]=(\5 )'(\4 )[\6 (\4 )]-(\6 )'(\4 )[\5 (\4 )]=
\hat {T}_N(\4 ),\label {pvfrelation}
\end{equation}
where $\hat {T}_N(\4 )$ is the perturbation vector field of the 
vector field
$ T=[K,S]\in T(M)$.
Therefore we have

\noindent (1) 
if $\sigma \in T(M)$ is an $n$-th order master-symmetry of the equation
$u_t=K(u)$, then $\hat{\sigma}_N=(\sigma ^{(0)T},\sigma ^{(1)T},\cdots,
\sigma ^{(N)T})^T\in T(\hat {M}_N)$ 
is an $n$-th order master-symmetry of the perturbation
equation $\hat {\eta}_{Nt}=\5 (\4 )$;

\noindent (2) the perturbation
equation $\hat{\eta }_{Nt}=\5 (\4 )$ possesses the same symmetry algebra 
structure as 
the original equation $u_t=K(u)$.
\end{theorem}
{\bf Proof:} By Lemma 2.2, we see for the $i$-th element that
\begin{eqnarray} &&(T(u))^{(i)}=
\bigl(K'(u)[S(u)]\bigr)^{(i)}(\hat {\1 }_i)-
\bigl(S'(u)[K(u)]\bigr)^{(i)}(\hat {\1 }_i)\nonumber\\ &=&
\bigl(\bigl(K(u)\bigr)^{(i)}\bigr)'(\hat {\1 }_i )[\hat {S }_i]-
\bigl(\bigl(S(u)\bigr)^{(i)}\bigr)'(\hat {\1 }_i )[\hat {K}_i]\nonumber\\
 &=&
\bigl(\5 '(\4 )[\hat {S}_N]\bigr)_i-\bigl(\6  '(\4 )[\hat {K}_N]\bigr)_i
,\ 0\le i\le N,\nonumber
\end{eqnarray} 
which shows (\ref{pvfrelation}) holds. The rest of the proof is obvious.
The proof is finished.
 \#

The relation (\ref{pvfrelation}) implies that the perturbation series 
(\ref{pseries}) keeps Lie product of vector fields invariant.

\begin{theorem} Let $K\in T(M)$.
When the evolution equation $u_t=K(u)$ has a Lax representation
$(L(u))_t=[A(u),L(u)]$ where $L,\,A$ are two matrix differential operators,
 the $N$-th order perturbation equation $\hat {\1 }_{Nt}=
\5 (\4 )$ has the following Lax representation
\begin{equation}
(\hat {L}_N(\4 ))_{t}=[\hat {A}_{N}(\4 ),\hat {L}_{N}(\4 )],\label{pelaxrep}
\end{equation}
where the spectral operator $\hat {L}_{N}$ and the Lax operator
$\hat {A}_{N}$ read as 
\begin{eqnarray}&&(\textrm{per}_NL)(\hat {\1 }_N)=
\hat {L}_{N}(\4 )\nonumber
\\
&=&\left[\bigl(\hat {L}_{N}(\4 )\bigr)_{ij}
\right]_{i,j=0,1,\cdots,N}=
\left[
\left.
\frac 1 {(i-j)!}\frac {\part ^{i-j}L (\3 )}{\part \2 ^{i-j}}
\right|_{\2 =0}
 \right]_{(N+1)\times(N+1)}\nonumber
\\
 &= &
\left[ \begin{array}{cccc}
L (\1 _0)& & &0\\ \left.
\frac {1}{1!}\frac {\part L (\3 )}{\part \2 }
\right|_{\2 =0}&L(\1 _0
) & 
& \\
\vdots & \ddots & \ddots & \\
 \left.
\frac {1}{N!}\frac {\part ^N L (\3 )}{\part \2 ^N}
\right|_{\2 =0}&\cdots &\left.
\frac {1}{1!}\frac {\part L (\3 )}{\part \2 }
\right|_{\2 =0}&L(\1 _0)
\end{array}
\right],
\end{eqnarray}
\begin{eqnarray}&&(\textrm{per}_NA)(\hat {\1 }_N)=
\hat {A}_{N}(\4 )\nonumber
\\
&=&\left[\bigl(\hat {A}_{N}(\4 )\bigr)_{ij}
\right]_{i,j=0,1,\cdots,N}=
\left[
\left.
\frac 1 {(i-j)!}\frac {\part ^{i-j}A (\3 )}{\part \2 ^{i-j}}
\right|_{\2 =0}
 \right]_{(N+1)\times(N+1)}\nonumber
\\
 &= &
\left[ \begin{array}{cccc}
A (\1 _0)& & & 0\\ \left.
\frac {1}{1!}\frac {\part A (\3 )}{\part \2 }
\right|_{\2 =0}&A(\1 _0
) & 
& \\
\vdots & \ddots & \ddots & \\
 \left.
\frac {1}{N!}\frac {\part ^N A (\3 )}{\part \2 ^N }
\right|_{\2 =0}&\cdots &\left.
\frac {1}{1!}\frac {\part A (\3 )}{\part \2 }
\right|_{\2 =0}&A(\1 _0)
\end{array}
\right].
\end{eqnarray}
\end{theorem}
{\bf Proof:} 
We first note that 
\[ \hat {u}_{Nt}=K(\hat{u}_N)+\textrm{o}(\2 ^N),\]
and thus we have
\begin{equation}
\left.\frac{\part ^{k}}{\part \2 ^{k}}\right|_{\2 =0}
\Bigl((L(\hat{u}_N))_t-[A(\hat{u}_N),L(\hat{u}_N)]\Bigr)=0,\ 0\le k\le N.
\label{laxrep}
\end{equation}
Let now us compute the elements of the differential operator 
matrix $[\hat {A}_N, \hat{L}_N]$.
Evidently, we know that   $[\hat {A}_N, \hat{L}_N]$ is lower-triangular, that is,
\[ ([\hat {A}_N, \hat{L}_N])_{ij}=0,\ 0\le i<j\le N.\]
On the other hand, when $0\le j\le i\le N$, 
we can compute
\begin{eqnarray}
(\hat {A}_N \hat{L}_N)_{ij}&=&\sum_{k=j}^i\frac{1}{(i-k)!}\left.\frac{
\part ^{i-k}A(\hat{u}_N)}{\part \2 ^{i-k}}\right|_{\2 =0}
\frac{1}{(k-j)!}\left.\frac{
\part ^{k-j}L(\hat{u}_N)}{\part \2 ^{k-j}}\right|_{\2 =0}\nonumber\\
&=& \frac 1{(i-j)!}\sum_{k=j}^i{ i-j \choose i-k}
\left.\frac{
\part ^{i-k}A(\hat{u}_N)}{\part \2 ^{i-k}}\right|_{\2 =0}
\left.\frac{
\part ^{k-j}L(\hat{u}_N)}{\part \2 ^{k-j}}\right|_{\2 =0}\nonumber
\\
&=& \frac 1{(i-j)!}\left.\frac{
\part ^{i-j}A(\hat{u}_N)L(\hat{u}_N)}{\part \2 ^{i-j}}\right|_{\2 =0}.\nonumber
\end{eqnarray}
In the same way, we obtain
\[(\hat {L}_N \hat{A}_N)_{ij}=
\frac 1{(i-j)!}\left.\frac{
\part ^{i-j}L(\hat{u}_N)A(\hat{u}_N)}{\part \2 ^{i-j}}\right|_{\2 =0}.
\]
Therefore we have
\[([\hat {A}_N, \hat{L}_N])_{ij}=
\frac 1{(i-j)!}\left.\frac{
\part ^{i-j}[A(\hat{u}_N),L(\hat{u}_N)]}{\part \2 ^{i-j}}\right|_{\2 =0},
\ 0\le i,j\le  N.\]
Now we easily find that (\ref{pelaxrep}) 
 is true due to (\ref{laxrep}). The proof is finished.
 \#

Note that the spectral operator $\hat {L}_N$ and the hereditary recursion
operator
$\hat {\Phi }_N$ have the same form of matrix. In fact, we can take 
the hereditary recursion operators as the spectral ones. More precisely,
$u_t=K(u)$ has a Lax representation$^{\cite{ChenLin}}$ $\Phi _t=[\Phi ,K']$ 
where $\Phi$ and $K'$ are a recursion operator and 
 the  Gateaux derivative operator of $K$, respectively.
The following result for the case of zero curvature 
representation may also be shown. Its proof is omitted due to the
completely 
similar deduction.

\begin{theorem} Let $K\in T(M)$.
When the evolution equation $u_t=K(u)$ has a zero curvature representation
$(U(u))_t-(V(u))_x+[U(u),V(u)]=0$
 where $U$ and $V$ are two matrix differential (sometimes multiplication)
operators,
 the $N$-th order perturbation equation $\hat {\1 }_{Nt}=
\5 (\4 )$ has the following zero curvature representation
\begin{equation}
(\hat {U}_N(\4 ))_{t}-(\hat {U}_N(\4 ))_x+
[\hat {U}_{N}(\4 ),\hat {V}_{N}(\4 )]=0,\label{pezcrep}
\end{equation}
where the two matrix differential operators $\hat {U}_{N}$ and
$\hat {V}_{N}$ are of  the form
\begin{eqnarray}&&(\textrm{per}_NU)(\hat {\1 }_N)=
\hat {U}_{N}(\4 )\nonumber
\\
&=&\left[\bigl(\hat {U}_{N}(\4 )\bigr)_{ij}
\right]_{i,j=0,1,\cdots,N}=
\left[
\left.
\frac 1 {(i-j)!}\frac {\part ^{i-j}U (\3 )}{\part \2 ^{i-j}}
\right|_{\2 =0}
 \right]_{(N+1)\times(N+1)}\nonumber
\\
 &= &
\left[ \begin{array}{cccc}
U (\1 _0)& & &0\vspace{1mm}\\ \left.
\frac {1}{1!}\frac {\part U (\3 )}{\part \2 }
\right|_{\2 =0}&U(\1 _0
) & 
&\vspace{1mm} \\
\vdots & \ddots & \ddots & \vspace{1mm}\\
 \left.
\frac {1}{N!}\frac {\part ^N U (\3 )}{\part \2 ^N}
\right|_{\2 =0}&\cdots &\left.
\frac {1}{1!}\frac {\part U (\3 )}{\part \2 }
\right|_{\2 =0}&U(\1 _0)
\end{array}
\right],
\end{eqnarray}
\begin{eqnarray}&&(\textrm{per}_NV)(\hat {\1 }_N)=
\hat {V}_{N}(\4 )\nonumber
\\
&=&\left[\bigl(\hat {V}_{N}(\4 )\bigr)_{ij}
\right]_{i,j=0,1,\cdots,N}=
\left[
\left.
\frac 1 {(i-j)!}\frac {\part ^{i-j}V (\3 )}{\part \2 ^{i-j}}
\right|_{\2 =0}
 \right]_{(N+1)\times(N+1)}\nonumber
\\
 &= &
\left[ \begin{array}{cccc}
V (\1 _0)& & &0\vspace{1mm}\\ \left.
\frac {1}{1!}\frac {\part V (\3 )}{\part \2 }
\right|_{\2 =0}&V(\1 _0
) & 
&\vspace{1mm} \\
\vdots & \ddots & \ddots &\vspace{1mm} \\
 \left.
\frac {1}{N!}\frac {\part ^N V (\3 )}{\part \2 ^N }
\right|_{\2 =0}&\cdots &\left.
\frac {1}{1!}\frac {\part V (\3 )}{\part \2 }
\right|_{\2 =0}&V(\1 _0)
\end{array}
\right].
\end{eqnarray}
\end{theorem}

\begin{theorem} Let $K\in T(M)$.
If the equation $u_t=K(u)$ possesses a Hamiltonian structure
\[ u_t=K(u)=J(u)\frac {\delta H(u)}{\delta u},\]
where $J:T^*(M)\to T(M)$ is a Hamiltonian operator and $H\in C^\infty (M)$
is a Hamiltonian function,
then the perturbation equation $\hat {\eta} _{Nt}=\5 (\4 )$ also possesses 
a Hamiltonian structure 
\begin{equation}
\hat {\eta} _{Nt}=\5 (\4 )=\7 (\4 )\frac {\delta \hat {H}_N(\4 )}{\delta \4 },
\end{equation} 
where the Hamiltonian operator $\7 (\4 )$ is given by (\ref{newham}) 
and the Hamiltonian function $\hat {H}_N\in C^\infty (\hat{M}_N)$
 is determined by
\begin{equation}
(\textrm{per}_NH)(\hat{\1 }_N)=
\hat {H}_N(\4 )= \frac 1{N!}\left. \frac {\part ^N H(\3 )}
{\part \2 ^N}\right |_{\2 =0}.\label{constantsformular}
\end{equation}
The corresponding Poisson bracket has the property
\be \{\hat {H}_{1N},\hat {H}_{2N}\}_{\hat {J}_{N}}(\hat{\1 }_N)
=\frac 1{N!}\left. \frac {\part ^N }
{\part \2 ^N}\right |_{\2 =0}\{H_1,H_2\}_J(\3 ),\ H_1,H_2\in C^\infty (M).
\label{Poissonpro}\ee
Furthermore the perturbation equation $\hat {\eta}_{Nt}=\5 (\4 )$
possesses a multi-Hamiltonian structure 
\[
\hat {\eta} _{Nt}=\5 (\4 )=
\hat{J}_{1N}(\4 )\frac{\delta \hat{H}_{1N}(\4 )}{\delta \4 }=
\cdots=
\hat{J}_{mN}(\4 )\frac{\delta \hat{H}_{mN}(\4 )}{\delta \4 },
\]
once $u_t=K(u)$ possesses an analogous
 multi-Hamiltonian structure
\[u_t=K(u)=J_1(u)\frac{\delta {H_1}(u )}{\delta u}=
\cdots=
J_m(u)\frac{\delta {H_m}(u )}{\delta u}.\]
\end{theorem}
{\bf Proof:} Let $\gamma =\frac {\delta H}{\delta u}\in T^*(M)$. Then we have
\begin{eqnarray}
&& \1 _{it}= \frac1 {i!}\left.\frac {\part ^i J(\3 )\gamma (\3 )}{\part \2 ^i}
\right|_{\2 =0}\nonumber \\
&=& \sum_{j=0}^i\frac 1{j!(i-j)!}
\left.\frac {\part ^{i-j} J(\3 )}{\part \2 ^{i-j}}
\right|_{\2 =0}\left.\frac {\part ^j \gamma (\3 )}{\part \2 ^j}
\right|_{\2 =0},\ 0\le i\le N.
\nonumber 
\end{eqnarray}
Thus we get
\begin{equation}
\hat {\eta}_{Nt}=\5 (\4 )=\7 (\hat {\eta}_N)\hat {\gamma }_N(\hat {\eta}_N),
\label{newhamiltonianstru}
\end{equation}
where the cotangent vector field $\hat {\gamma }_N\in T^*(\hat {M}_N)$ reads as
\be \hat {\gamma }_N(\hat {\eta}_N)=
\Bigl( \frac 1 {N!}\left.\frac 
{\part ^N\gamma ^T(\3 )}{\part \2 ^N}\right|_{\2 =0}
,\frac 1 {(N-1)!}\left.\frac {\part ^{N-1}\gamma ^T(\3 )}{\part \2 ^{N-1}}
\right|_{\2 =0}
, \cdots, \frac 1 {1!}\left.\frac {\part \gamma ^T(\3 )}{\part \2 }\right|_{\2 =0}
, \gamma ^T(\eta _0)\Bigr)^T.\label{pcvf}
\ee
We hope that the cotangent vector field
 $\hat {\gamma }_N$ is a gradient field. If so,
the potential function should be the following
\begin{eqnarray} &&\hat {H}_N(\4 )
=\int _0^1<\hat {\gamma }_N(\lambda \4 ), \4 >\,
d\lambda \nonumber\\&=&
\int_0^1\sum_{i=0}^N\frac 1{i!}<\left.\frac {\part ^i\gamma (\lambda
\3 )}{\part \2 ^i}\right|_{\2 =0},\eta _{N-i}>\, d\lambda\nonumber
\\&=&\frac 1 {N!}\left.\frac {\part ^N}{\part \2 ^N}\right|_{\2 =0}
\int _0^1<\gamma (\lambda \3 ), \3 >\,
d\lambda =\frac 1 {N!}\left.\frac
 {\part ^N H(\3 )}{\part \2 ^N}\right|_{\2 =0}.
\nonumber 
\end{eqnarray}
Actually, the cotangent 
vector field $\hat {\gamma }_N$ is a gradient field.
We can show that 
\begin{equation}
\hat {\gamma }_N(\4 )=\frac {\delta \hat {H}_N(\4 )}{\delta \4 }.
\label{pvector}\end{equation}
According to the definition of the variational derivative, we have for any 
$S_i\in T(M(\eta _i))$
\begin{eqnarray} &&<\frac {\delta }{\delta \1 _i }\,\Bigl( \frac 1 {N!}
\left.\frac {\part ^N H(\3 )}{\part \2 ^N}\right|_{\2 =0}
\Bigr), S_i(\eta_i) >=\Bigl( \frac 1 {N!}
\left.\frac {\part ^N H(\3 )}{\part \2 ^N}\right|_{\2 =0}
\Bigr)'(\eta _i)[S_i(\eta_i) ]\nonumber\\&=&
\frac 1 {N!}
\left.\frac {\part ^N }{\part \2 ^N}\right|_{\2 =0}
H'(\3 )[\2 ^iS_i(\eta_i) ]=\frac 1 {(N-i)!}
\left.\frac {\part ^{N-i} }{\part \2 ^{N-i}}
\right|_{\2 =0}H'(\3 )[S_i(\eta_i) ]
\nonumber\\
&=&\frac 1 {(N-i)!}
\left.\frac {\part ^{N-i} }{\part \2 ^{N-i}}\right|_{\2 =0}<\frac 
{\delta H(\3 )}{\delta \3 }, S_i(\eta_i) >=
\frac 1 {(N-i)!}
\left.\frac {\part ^{N-i} }
{\part \2 ^{N-i}}\right|_{\2 =0}<\gamma (\3 ), S_i(\eta_i) >
\nonumber\\
&=&<\frac 1 {(N-i)!}
\left.\frac {\part ^{N-i}\gamma (\3 ) }
{\part \2 ^{N-i}}\right|_{\2 =0}, S_i(\eta_i) >,\ 0\le i\le N.
\nonumber 
\end{eqnarray}
This shows that (\ref{pvector}) holds, indeed. Therefore 
(\ref{newhamiltonianstru}) is a Hamiltonian equation.

Let us now prove the property (\ref{Poissonpro}). Let $\beta =
\frac {\delta H_{1}}{\delta u},\gamma =\frac {\delta H_{2}}{\delta u}
\in T^*(M)$. In virtue of (\ref{pvector}), we can compute that
\begin{eqnarray}&& \{\hat{H}_{1N},\hat{H}_{2N}\}_{\hat{J}_{N}}(\hat{\eta}_{N})
=<\frac {\delta \hat{H}_{1N}(\hat{\eta}_{N})}{\delta \hat{\eta}_{N}},
\hat{J}_{N}(\hat{\eta}_{N})
\frac {\delta \hat{H}_{2N}(\hat{\eta}_{N})}{\delta \hat{\eta}_{N}}>
\nonumber\\&=&
\sum_{i=0}^N<\frac 1 {(N-i)!}\left.
\frac {\part ^{N-i}\beta (\hat {u}_N)}{\part \2 ^
{N-i}}\right|_{\2 =0},\sum_{j=N-i}^N\frac 1 {(i+j-N)!}\times
\nonumber\\&&\qquad\quad
 \left.
\frac {\part ^{i+j-N}J(\hat {u}_N)}{\part \2 ^
{i+j-N}}\right|_{\2 =0}\frac 1 {(N-j)!}
\left.\frac {\part ^{N-j}\gamma (\hat {u}_N)}{\part \2 ^
{N-j}}\right|_{\2 =0}>
\nonumber\\&=&
\sum_{i=0}^N<\frac 1 {(N-i)!}\left.
\frac {\part ^{N-i}\beta (\hat {u}_N)}{\part \2 ^
{N-i}}\right|_{\2 =0}, \frac 1 {i!}\left.
\frac {\part ^{i}(J (\hat {u}_N)\gamma (\hat {u}_N))}{\part \2 ^
{i}}\right|_{\2 =0}>
\nonumber\\&=&
\frac 1{N!}\left.\frac 
{\part ^N}{\part \2 ^N}\right|_{\2 =0}
<\beta (\hat {u}_N),J(\hat {u}_N)
\gamma (\hat {u}_N)>=\frac 1{N!}\left.\frac 
{\part ^N}{\part \2 ^N}\right|_{\2 =0}
\{H_1,H_2\}_J(\hat {u}_N).
\nonumber
\end{eqnarray}
It follows that the equality (\ref{Poissonpro}) is true.

Further noting the concrete form of new Hamiltonian operators,
a multi-Hamiltonian structure may readily be 
established for the perturbation equation.
The proof is completed.
\#
\label{theory}

We should note that two important formulas: (\ref{constantsformular}) and
(\ref{pcvf}). (\ref{constantsformular}) provides a explicit formula
for computing constants of motion of the $N$-th perturbation equations
and (\ref{pcvf}) gives rise to an expression of perturbation cotangent vector 
fields.

\section{Applications to integrable equations}
\setcounter{equation}{0}
\subsection{MKdV hierarchy}

We first consider the following MKdV hierarchy$^{\cite{Maphysicaa}}$
\be u_{t_n}=K_n=a_{nx}=J\Psi ^na_0=J\frac {\delta H_n}{\delta u},\ n\ge 0,
\label{mkdvh}
\ee 
with \[
J=\part , \ \Psi(u)=-\frac14 \part ^2+u\part ^{-1}u\part ,\ 
H_n(u)=\frac {b_{n+1}-
c_{n+1}}{2(2n+1)},\ n\ge 0,\]
where $a_i,\,b_i\,c_i,\ i\ge0,$ are recursively defined by
 \[\left \{\begin{array}{l}
a_0=u,\ b_0=1,\ c_0=-1,\vspace{0.5mm}\\
a_{i+1}=\Psi a_i, 
\ i\ge0,\vspace{0.5mm} \\ 
b_{i+1}=\frac12
a_{ix}+\part ^{-1}u\part a_i,\ i\ge0,\vspace{0.5mm}\\
c_{i+1}=\frac12a_{ix}-\part ^{-1}u\part a_{i},\ i\ge0,\end{array}
\right.\]
and $\part \part ^{-1}=\part ^{-1}\part=1.$
The first equation is exactly MKdV equation
 \be u_{t_1}=-\frac14 u_{xxx}+\frac 32u^2u_x.\label{mkdv}\ee
Its inverse scattering transform was first studied by 
Wadati$^{\cite{Wadati}}$.
The MKdV hierarchy (\ref{mkdvh})
possesses zero curvature representations$^{\cite{Maphysicaa}}$
\be U_{t_n}-V^{(n)}_x+[U,V^{(n)}]=0,\ n\ge0,\ee 
with 
\be U=\left[\ba {cc}u&-\lambda \vspace{1mm}\\ -1&-u\ea  \right],\ 
 V^{(n)}=\left[\ba {cc}(a\la ^n)_+&(b\la ^n)_+\lambda 
\vspace{1mm}\\(c\la ^n)_+ 
&-(a\la ^n)_+\ea  \right],\ n\ge 0.
\ee 
Here the plus symbol $+$ stands for the choice of non-negative power of $\la $
and \[
a=\sum_{i=0}^\infty a_i\la ^{-i},\ b=\sum_{i=0}^\infty b_i\la ^{-i},\ 
c=\sum_{i=0}^\infty c_i\la ^{-i}.\]

Applying the integrable 
theory in Sec. \ref{theory}, 
we can obtain  infinitely many new
hereditary operators 
$\textrm{per}_{N_1}\textrm{per}_{N_2}\cdots\textrm{per}_{N_m}\Phi$,
$N_1,\,N_2\,\cdots,\, N_m\ge1$, starting
from the hereditary operator of MKdV hierarchy
\[\Phi =\Psi ^\dagger =-\frac14 \part ^2+\part u\part ^{-1}u.\] 
We easily find that some explicit expressions:
\begin{eqnarray}&& \ (\textrm{per}_N\Phi)(\hat {\eta} _N)
 =\textrm{diag}(\,\,
\underbrace{-\frac14\part ^2,\cdots,-\frac14\part ^2}_{N+1}\,
\,)+
\left[ \sum_{k=0}^{i-j}\part \eta _k\part ^{-1}\eta _{i-j-k}\right]_{(N+1)
\times (N+1)}\nonumber \\ &&
=\left[\ba {cccc}
-\frac14 \part ^2 +\part \eta _0\part ^{-1}\eta _0& 
 & & 0\vspace{2mm}\\
\part \eta _0\part ^{-1}\eta _1+\part \eta _1\part ^{-1}\eta _0&\ddots
&  & \vspace{2mm}\\
\vdots & \ddots & \ddots & \vspace{2mm}\\
\sum_{k=0}^N \part \eta _k\part ^{-1}\eta _{N-k}&\cdots &
\part \eta _0\part ^{-1}\eta _1+\part \eta _1\part ^{-1}\eta _0&
-\frac14 \part ^2 +\part \eta _0\part ^{-1}\eta _0\ea
 \right],\nonumber 
\end{eqnarray}
\[\ba{l} \ 
(\textrm{per}_{1}\textrm{per}_{1}
\Phi)(\eta ) =   \vspace{2mm}\\
\left[
\ba {cccc} -\frac14 \part ^2 +\part \eta _0\part ^{-1}\eta _0& 
0 &0 & 0\vspace{1mm}\\
\part \eta _0\part ^{-1}\eta _1+\part \eta _1\part ^{-1}\eta _0&
-\frac14 \part ^2 +\part \eta _0\part ^{-1}\eta _0&0&0\vspace{1mm}\\
\part \eta _0\part ^{-1}\eta _2+\part \eta _2\part ^{-1}\eta _0&0&
-\frac14 \part ^2 +\part \eta _0\part ^{-1}\eta _0&0\vspace{1mm}\\
\sum_{k=0}^3
\part \eta _k\part ^{-1}\eta _{3-k}&
\part \eta _0\part ^{-1}\eta _2+\part \eta _2\part ^{-1}\eta _0&
\part \eta _0\part ^{-1}\eta _1+\part \eta _1\part ^{-1}\eta _0&
-\frac14 \part ^2 +\part \eta _0\part ^{-1}\eta _0
\ea \right]
\ea\]
where $\eta =(\eta _0,\eta_1,\eta _2,\eta _3)^T.$ It needs a large amount
of calculation if we directly prove the hereditariness of 
the above two operators.

The $N$-th order perturbation equation of $u_{t_n}=K_n$  reads as
\be 
\hat{\eta }_{Nt_n}=
(\textrm{per}_{N}K_n)(\hat{\1 }_N)=(\textrm{per}_{N}\Phi)(\hat{\1 }_N)(
 \textrm{per}_{N}K_{n-1})(\hat{\1 }_N)
=((\textrm{per}_{N}\Phi)(\hat{\1 }_N))^n\hat{\eta }_{Nx},\label{pmkdvh}
\ee
among which is the $N$-th order perturbation equation
of MKdV equation (\ref{mkdv}) 
\begin{eqnarray}  \eta _{it_1}&=&
-\frac14\part ^2\eta _{ix}+\sum_{l=0}^i\sum_{k=0}^l
\part \eta _k\part ^{-1}\eta _{l-k}\eta_{i-l,x}\nonumber
\\&=&
-\frac14 \1 _{ixxx}+\frac32 \sum_{\textrm{\scriptsize{$\ba {c}
 j+k+l=i\\0\le j,k,l\le i\ea $}}}
\1 _{jx}\1 _k\1 _l
,\
0\le i\le N.\label{pmkdv}
\end{eqnarray}
The resulting local evolution equations defined by (\ref{pmkdvh}) are
all integrable soliton ones for any $n,N\ge 1$.
They all have zero curvature representations 
\[((\textrm{per}_{N}U)(\hat{\1 }_N))_{t_n}-((\textrm{per}_{N}V^{(n)})
(\hat{\1 }_N))_{x}+
\left[\right.(\textrm{per}_{N}U)(\hat{\1 }_N),(
 \textrm{per}_{N}V^{(n)})(\hat{\1 }_N)\left.\right]=0,\]
and bi-Hamiltonian formulations 
\[ \1 _{Nt_n}=(
 \textrm{per}_{N}J)(\hat{\1 }_N)\frac {\delta (
 \textrm{per}_{N}H_n)(\hat{\1 }_N)}{\delta \hat{\eta}_N}=(
 \textrm{per}_{N}M)(\hat{\1 }_N)\frac {\delta (
 \textrm{per}_{N}H_{n-1})(\hat{\1 }_N)}{\delta \hat{\eta}_N}\]
where the Hamiltonian operator $M=JL=-\frac14 \part ^3+\part u\part ^{-1}u
 \part $. Therefore they possess infinitely many common symmetries
$\textrm{per}_{N}K_m,\ m\ge0$, and infinitely many common constants of 
motion $\textrm{per}_{N}H_m,\ m\ge0$. In particular,
for the $N$-th perturbation equation (\ref{pmkdv}) of 
MKdV equation (\ref{mkdv})
we can get the following explicit results:
two Hamiltonian functions
\begin{eqnarray}&&
(\textrm{per}_{N}H_0)(\hat{\1 }_N)= \frac12 \sum_{j=0}^N\1 _j\1 _{N-j}
,\nonumber \\ &&
(\textrm{per}_{N}H_1)(\hat{\1 }_N)=
-\frac1 {12} \sum_{j=0}^N \1 _j\1 _{N-j,xx}+\frac 1 {24}
\sum_{j=0}^N \1 _{jx}\1 _{N-j,x}\nonumber\\ && \qquad\ 
\qquad\qquad +\frac1 8 \sum_{\textrm{{\scriptsize $ \ba {c} 
j_1+j_2+j_3+j_4=N\\
0\le j_1,j_2,j_3,j_4\le N
\ea 
$}}}\1 _{j_1}\1 _{j_2}\1 _{j_3}\1 _{j_4}\nonumber
 ;\end{eqnarray}
a Hamiltonian pair
\[ (\textrm{per}_{N}J)(\hat{\1 }_N)=\textrm{diag}(\, \underbrace{\part ,
\part ,\cdots, \part }_{N+1}\, ),\]
\[(\textrm{per}_{N}M)(\hat{\1 }_N)=
\left[\ba {cccc}  0 & & &
-\frac14 \part ^3 +\part \1 _0\part ^{-1}\1 _0 \part \vspace{2mm}\\
 & &\vdots & \sum_{k=0}^1\part \1 _k\part ^{-1}\1 _{1-k} \part \vspace{2mm}\\
 &\vdots &\vdots &\vdots\vspace{2mm}\\
-\frac14 \part ^3 +\part \1 _0\part ^{-1}\1 _0 \part &
\sum_{k=0}^1\part \1 _k\part ^{-1}\1 _{1-k} \part &
\cdots &\sum_{k=0}^N\part \1 _k\part ^{-1}\1 _{N-k} \part
\ea
\right];
\]
and a pair of Lax operators
\[(\textrm{per}_{N}U)(\hat{\1 }_N)=
\left[\ba {cccc}A_0& & &0\\A_1&A_0& & \\ \vdots &\ddots&\ddots& 
\\A_N&\cdots&A_1&A_0\ea \right]
 ,\ 
(\textrm{per}_{N}V^{(1)})(\hat{\1 }_N)=
\left[\ba {cccc}
B_0& & &0\\B
_1&B_0& & \\\vdots &\ddots&\ddots& 
\\B_N&\cdots&B_1&B_0\ea \right]
 ,\]
where the operators $A_i,\ B_i,\ 0\le i\le  N,$ are given by
\[\ba {l}A_i=\left[\ba {cc}\1 _i &\la \\ -1 & -\1 _i
 \ea\right],\ \  B_i=  \vspace{2mm} \\ 
\left[\ba {cc} 
\1 _i\la -\frac14 \1 _{ixx}+\frac12 \sum_{\textrm{{\scriptsize $ 
\ba {c}j_1+j_2+j_3=i\\0\le j_1,j_2,j_3\le i \ea $}}}\1 _{j_1}\1 _{j_2}\1 _{j_3}
&\delta _{i0}\la ^2+\frac12 (\1 _{ix}+\sum_{j=0}^i\1 _j\1 _{i-j})\la 
\vspace{2mm}\\
-\delta_{i0}\la +\frac12 (\1 _{ix}-\sum_{j=0}^i\1 _j\1 _{i-j})&
-\1 _i\la +\frac14 \1 _{ixx}-\frac12 
\sum_{\textrm{{\scriptsize $ 
\ba {c}j_1+j_2+j_3=i\\0\le j_1,j_2,j_3\le i \ea $}}}\1 _{j_1}\1 _{j_2}\1 _{j_3}
\ea\right].
\ea \]
In addition, we can generate a $\tau$-symmetry algebra$^{\cite{ma1990jpa}}$
of the perturbation equation (\ref{pmkdvh}) by a perturbation of
 $\tau$-symmetries of $u_{t_n}=K_n$. Moreover we may also consider 
the nonlinear problem of the Lax systems 
\[\hat{\phi }_{Nx}=(\textrm{per}_NU)(\hat{\eta} _N)\hat{\phi }_{N},\ 
\hat{\phi }_{Nt_n}=(\textrm{per}_NV^{(n)})(\hat{\eta} _N)\hat{\phi }_{N},\] 
similar to Ref. \cite{Maphysicaa}.

\subsection{Kadomtsev-Petviashvili equation}
Let us now consider the Kadomtsev-Petviashvili equation
\be  u_t=K(u)=K_3(u)=\part ^{-1}_xu_{yy} -u_{xxx}-6uu_x.\label{kp}\ee 
It has Lax pair
\be L(u)=\frac i{\sqrt{3}}\part _y +\part ^2_x +u,\ A(u)=3i\part ^{-1}_xu_y
-3u_x-6u\part _x -4\part ^3_x,\ i=\sqrt{-1}
\ee 
and time independent symmetries
\[ K_1=\frac32 u_x,\  K_2=u_y,\ K_n=[K_{n-1},\tau ],\ n\ge2,\]
where $\tau =yK_3+\frac23 xu_y+\frac43\part _x^{-1}u_y$
and further the $k$-th order master symmetry (see 
for example Ref. \cite{ma1})
\[ \tau^{(k)}=y^{k-1},\ \tau ^{(k)}_{i_1i_2\cdots i_j}=
\textrm{ad}_{K_{i_1}}\textrm{ad}_{K_{i_2}}\cdots\textrm{ad}_{K_{i_j}}
y^{k+j-1},\ j,k\ge1 .\]  
Therefore KP equation (\ref{kp}) possesses the following time 
polynomial dependent symmetries
\[
\sigma ^{(k)}=\sum_{l=0}^k\frac {t^l}{l!}(\textrm{ad}_K)^l \tau ^{(k)},\
\sigma ^{(k)}_{i_1i_2\cdots i_j}
=\sum_{l=0}^k\frac {t^l}{l!}(\textrm{ad}_K)^l \tau ^{(k)}
_{i_1i_2\cdots i_j},\ j,k\ge1,
\]
which contain the symmetries with the forms $\frac {\part }{\part t}
\sigma ^{(k)}$ and  $\frac {\part }{\part t}
\sigma ^{(k)}_{i_1i_2\cdots i_j}$.

The corresponding perturbation equation
$\hat{\eta} _{Nt}=\hat{K} _N$ becomes
\be \eta _{it}=\part ^{-1}_x\eta _{iyy}-\eta _{ixxx}-6\sum _{j=0}^i
\eta_j\eta _{i-j,x},\ 0\le i\le N.\label{pkp}
\ee
Following the general theory in Sec. \ref{integrableproperties},
(\ref{pkp}) is also an integrable equation in $2+1$ dimensions.
For example, it has Lax pair
$ (\hat {L}_n(\4 ))_{t}=[\hat {A}_{N}(\4 ),\hat {L}_{N}(\4 )]
$,
where the spectral operator $\hat {L}_{N}$ and the Lax operator
$\hat {A}_{N}$ read as
\[ \hat {L}_{N}=
\left[ \begin{array}{cccc}
\frac i{\sqrt{3}}\part _y +\part ^2_x +\1 _0& & 0\\ 
\eta _1
&\frac i{\sqrt{3}}\part _y +\part ^2_x +\1 _0
 & 
& \vspace{3mm}\\
\vdots & \ddots & \ddots &\vspace{3mm} \\
\eta _N
&\cdots &\eta _1
&\frac i{\sqrt{3}}\part _y +\part ^2_x +\1 _0
\end{array}
\right],
\]
\[\hat {A}_{N}
=
\left[ \begin{array}{cccc}
A(\eta _0)
& & 0\\ 3i\part ^{-1}_x\eta _{1y}
-3\eta _{1x}-6\eta _1\part _x 
&A(\1 _0
) & 
& \vspace{3mm}\\
\vdots & \ddots & \ddots & \vspace{3mm}\\
3i\part ^{-1}_x\eta _{Ny}
-3\eta _{Nx}-6\eta _N\part _x 
&\cdots &
3i\part ^{-1}_x\eta _{1y}
-3\eta _{1x}-6\eta _1\part _x 
&A(\1 _0)
\end{array}
\right].
\]
Moreover (\ref{pkp}) has the $k$-th order master symmetry 
\[ \textrm{per}_N\tau ^{(k)}=(\,
y^{k-1},\underbrace{0,\cdots,0}_N\,)^T,\ k\ge1,\]
and thus possesses time polynomial dependent symmetries
\[ \textrm{per}_N\sigma ^{(k)} =\sum _{l=0}^k\frac {t^l}{l!} (\textrm{ad}_{
\textrm{per}_NK})^l\textrm{per}_N\tau ^{(k)},\ k\ge1,
\]
and \[
\textrm{per}_N\sigma ^{(k)}_{i_1i_2\cdots i_j}
 =\sum _{l=0}^k\frac {t^l}{l!} (\textrm{ad}_{
\textrm{per}_NK})^l\textrm{per}_N\tau ^{(k)}_{i_1i_2\cdots i_j},\ j,k\ge1.
\]
Although KP equation has no regular recursion operator$^{\cite{ZakharovK}}$,
we may also construct a bi-Hamiltonian formulation of (\ref{pkp}) through
the perturbation of the bi-Hamiltonian one$^{\cite{DorfmanF}}$. But here we 
omitted the discussion because of the complicated notation.

Let us now take another form of the  Kadomtsev-Petviashvili
equation 
\be  u_{xt}
=\part _x K=u_{yy}-(u_{xxx}+6uu_x)_x.\label{gkp}\ee 
Evidently its solutions include all solutions of the Kadomtsev-Petviashvili
equation (\ref{kp}) and hence 
it is a little more general than (\ref{kp}). 
We are about to see that there exists a different symmetry property 
between (\ref{kp}) and (\ref{gkp}). 
Now the corresponding linearized equation to (\ref{gkp})
is as follows
\be  \sigma _{xt}=\part _x (K'[\sigma ])=\part _{yy}\sigma -\part _x^4\sigma
-6\part _x^2(u\sigma )  .\label{lgkp}\ee
One may directly prove that (\ref{gkp}) possesses the following three time
 dependent
symmetries
\begin{eqnarray}  &&
\sigma ^{(1)}(f)=-\frac 16f_t+fu_x,\label{sigma1}\\&&
\sigma ^{(2)}(f)=-\frac1 {12}f_{tt}y+\frac12 f_tyu_x+fu_y,\label{sigma2}\\&&
\sigma ^{(3)}(f)=-\frac1 {36}f_{ttt}y^2+f_{tt}(\frac16 y^2u_x-\frac1 {18}
x)
\nonumber\\&&
\qquad\qquad 
+f_t(\frac 23 u+\frac23 yu_y +\frac13 xu_x)+fu_t,\label{sigma3}
\end{eqnarray} 
with an arbitrary function $f$ of $t$, 
 each of which is not any symmetry of (\ref{kp}) while $
f_{tt}\ne 0,\ f_{ttt}\ne0$ or $ f_{tttt}\ne0$ respectively.
That kind of symmetries
is first introduced in Ref. 
\cite{Schwarz}. In fact, there is a rule to generate 
these symmetries. For example, 
\[\sigma ^{(2)}(f)=-\frac1 {12} f_{tt}y+f_t[K,-\frac1 {12}y]+
f[K,[K,-\frac1 {12}y]].\]
We can also construct new vector fields 
\be\sigma ^{(n)}(f)
=\sum_{i=0}^n\frac {d^{n-i}f}{dt^{n-i}}(\textrm{ad}_K)^iy^{n-1}
,\ n\ge4.\label{newvf}\ee
But in these vector fields, there exist nonlinear terms involving $u_y$ or
 $u$ etc., which can't be balanced in the linearized equation
(\ref{lgkp}). Therefore (\ref{newvf}) are not any symmetries of 
(\ref{gkp}). The symmetries determined by (\ref{sigma1}), (\ref{sigma2})
and (\ref{sigma3})
constitute a Lie algebra
\be \left\{\begin{array}{l} \left [\right.\sigma ^{(1)}(f),\sigma ^{(1)}(g)
\left.\right
]=0,\vspace{2mm}
\\ \left
[\right.\sigma ^{(2)}(f),\sigma ^{(2)}(g)\left.\right 
]=\sigma ^{(1)}(\frac12 fg_t-\frac12 f_tg),
\vspace{2mm}\\ \left
[\right.\sigma ^{(3)}(f),\sigma ^{(3)}(g)\left.\right
 ]=\sigma ^{(3)}(fg_t-f_tg),\vspace{2mm}\\
\left[\right.\sigma ^{(1)}(f),\sigma ^{(2)}(g)\left.\right ]=0,\vspace{2mm}\\
\left[\right.\sigma ^{(1)}(f),\sigma ^{(3)}(g)\left.\right 
]=\sigma ^{(1)}(\frac13 fg_t- f_tg),
\vspace{2mm}\\
\left [\right.\sigma ^{(2)}(f),\sigma ^{(3)}(g)\left.\right
]=\sigma ^{(2)}(\frac23 fg_t- f_tg),
\end{array}\right. \label{symmalg} \ee 
where $f,g$ are two functions of $t$. Note that the second commutator
relation is somewhat different from one given in Ref. \cite{DavidKLW}.
But if we choose $\tilde {\sigma }^{(2)}(f)=\sqrt{2}{\sigma }^{(2)}(f)$,
then two algebras are the same.
The above symmetry algebra contains 
the following three subalgebras 
$\{\sigma^{(1)}(f), \ \sigma^{(2)}(f)\}$, $\{\sigma^{(3)}(f)\}$ and
$\{\sigma^{(1)}(f), \ \sigma^{(2)}(f),\ \sigma^{(3)}(1)=u_t\}$.
The last subalgebra has the simple commutator relations
\[ [u_t,\sigma^{(1)}(f)]=\sigma^{(1)}(f_t),\ 
[u_t,\sigma^{(2)}(f)]=\sigma^{(2)}(f_t).
\]

It is easy to see that
the $N$-th order perturbation equation of (\ref{gkp}) 
\be \eta _{ixt}=\eta _{iyy}-(\eta _{ixxx}+6\sum _{j=0}^i
\eta_j\eta _{i-j,x})_x,\ 0\le i\le N\label{pgkp}
\ee
has also the similar symmetry
property. 
More precisely, it possesses the following time dependent symmetries
with an arbitrary function of $t$
\[\begin{array}{l}\textrm{per}_N\sigma ^{(1)}(f)=(-\frac16 f_t+f
\eta _{0x},f\eta _{1x},
\cdots,f\eta _{Nx})^T,\vspace{2mm}\\
\textrm{per}_N\sigma ^{(2)}(f)=(-\frac1{12} f_{tt}y+\frac12
f_ty\eta _{0x}+f\eta _{0y},\frac12
f_ty\eta _{1x}+f\eta _{1y},\cdots,\frac12
f_ty\eta _{Nx}+f\eta _{Ny})^T,\vspace{2mm}\\
\textrm{per}_N\sigma ^{(3)}(f)=(
-\frac1 {36}f_{ttt}y^2+f_{tt}(\frac16 y^2\eta_{0x}-\frac1 {18}
x)+f_t(\frac 23 \eta_0+\frac23 y\eta_{0y}
 +\frac13 x\eta_{0x})+f\eta_{0t},\vspace{2mm}\\
\qquad\qquad\qquad f_{tt}(\frac16 y^2\eta_{1x}-\frac1 {18}
x)+f_t(\frac 23 \eta_1+\frac23 y\eta_{1y}
 +\frac13 x\eta_{1x})+f\eta_{1t},\vspace{2mm}\\
\qquad\qquad\qquad \cdots,f_{tt}(\frac16 y^2\eta_{Nx}-\frac1 {18}
x)+f_t(\frac 23 \eta_N 
+\frac23 y\eta_{Ny} +\frac13 x\eta_{Nx})+f\eta_{Nt})^T,\vspace{2mm}
\end{array}\]
which constitute the same infinite dimensional symmetry algebra as
(\ref{symmalg}).
In fact, the equation (\ref{pgkp})
has the same integrable property as (\ref{gkp}). 
It is an interesting problem how to construct integrable equations 
which possess a kind of symmetries involving an arbitrary function 
of time variable. 

It is well known that the evolution equation has not a similar property,
i.e. it doesn't possess$^{\cite{Ma2}}$ the following 
symmetries involving an arbitrary function $f$
of the time variable
\[ \sigma ^{(n)}=\sum_{i=0}^n\frac {d^{n-i}f}{dt^{n-i}}S_i(u),\]
where $S_i,\ 0\le i\le n,$ don't depend explicitly on the time variable.
In general, it possesses time polynomial
dependent symmetries generated by its master 
symmetries$^{\cite{Fuchssteiner1983}}$. Therefore (\ref{kp}) and (\ref{gkp}) 
have different symmetry algebras.

\section{Concluding remarks}
\setcounter{equation}{0}

We may also make another perturbation series 
\be
\check {u}_N 
=\sum_{i=0}^N\frac {\2 ^i}{i!}\xi  _i,\label{pseries2}\ee
similar to the perturbation series (\ref{pseries}).
This moment, the corresponding perturbation equation reads as
\begin{equation}
\check {\xi  }_{Nt} = \check {K}_N (\check {\xi  }_N),\ 
\check {\xi  }_N=(\xi  _0^T,\xi  _1^T,\cdots,\xi  _N^T)^T, \label{checkpe}
\end{equation} 
where the perturbation vector field is of the form
\[
 \check {K}_N (\check {\xi  }_N)=\Bigl(K^T(\xi  _0),
\left.\frac {\part K^T (\check {u}_N  )}{\part \2 }
\right|_{\2 =0},\cdots,\left.\frac {\part ^N K ^T
(\check {u}_N  )}{\part \2 ^N }
\right|_{\2 =0}\Bigr)^T.\]
In an analogous way, we can generate another new hereditary operator
\begin{equation}
\check {\Phi}_N(\check {\xi  }_N)=
\left[
\left.
{i \choose j}
\frac {\part ^{i-j}\Phi (\check {u}_N )}{\part \2 ^{i-j}}
\right|_{\2 =0}
 \right]_{i,j=0,1,\cdots,N}
\end{equation}
from  a known 
hereditary operator $\Phi (u)$, and $\check {\Phi}_N(\check {\xi  }_N)$
is a recursion operator of (\ref{checkpe}) provided that $\Phi (u)$ is a
 recursion operator of
the original equation $u_t=K(u)$. 
From the above expression, we see that the formation of 
the perturbation equation under the perturbation series (\ref{pseries2})
is simpler but the corresponding recursion operator 
is more complicated than ones under (\ref{pseries}). 
However there exists an intimate
 relation between two kinds of
 perturbations because we have a Miura transformation
\be \1 _i=\1 _i (\xi  _0,\xi  _1,\cdots,\xi  _N
 )=\frac 1{i!}\xi  _i,\ 0\le i\le N.\label{miurat}\ee
 For example, 
the operator $\check {\Phi}_N(\check {\xi  }_N)$ may be generated
by a transformation 
\be \ba {l}\ \ 
 \check {\Phi}_N(\check {\xi  }_N)=\frac {\part \check \xi _N}
{\part \hat \eta _N}
\hat {\Phi }_N(\hat {\eta }_N)\left(\frac {\part \check \xi _N}
{\part \hat \eta _N}\right)^{-1}\vspace{2mm}\\=
\left[\begin{array}{ccccc}1& & & &0\\
 & {1!}& & & \\
 & & {2!}& & \\
 & & &\ddots& \\
0 & & & & {N!}\end{array}\right]
\hat {\Phi}_N(\hat {\eta }_N)
\left[\begin{array}{ccccc}1& & & &0\\
 &{1!}& & & \\
 & & {2!}& & \\
 & & &\ddots& \\
0 & & & & {N!}\end{array}\right]^{-1}.\ea \ee
In fact, we can similarly obtain any new tensors by means of 
the Miura transformation (\ref{miurat}). For instance,
a new Hamiltonian operator may be engendered by
\be  \check J _N(\check \xi _N)=\frac {\part \check \xi _N}
{\part \hat \eta _N}
\hat {J }_N(\hat {\eta }_N)\left(\frac {\part \check \xi _N}
{\part \hat \eta _N}\right)^\dagger 
=\left[\left. \frac {i!j!}{(i+j-N)!}\frac {\part ^{i+j-N} 
J(\check u _N)}{\part \2 ^{i+j-N}}\right|_{\2 =0}\right]_{i,j=0,1,\cdots,N}.
\ee

We remark that 
by the resulting perturbation equations in Sec. \ref{theory},
 we can generate approximate solutions of 
the original equations to a precision $o(\2 ^N)$. This is different from
the construction of the $\tau $ functions in bilinear formation, 
where the expansion series holds exactly for any order precision.
It is also of interest to note that the perturbation equations
are all integrable coupling with the original equations and the original ones
always appear in the first position. Therefore our integrable theory provides 
an approach for constructing integrable coupling
of soliton equations and
enriches the intention of perturbation bundle established in Ref. 
\cite{Fuchssteiner1}. However 
it is still a problem deserving of investigation
how to construct more general 
integrable coupling by perturbation. 


\noindent{\bf Acknowledgments:} One of the authors (W. X. Ma) 
would like to thank the 
Alexander von Humboldt Foundation,
the National Natural 
Science Foundation of China and the Shanghai Science and Technology Commission
of China
 for their financial support. 
He has benefited from helpful discussions with Dr. 
W. Oevel.

\setlength{\baselineskip}{13.3pt}

\end{document}